\documentclass[11pt]{article}
\usepackage{utils}

\begin{document}

\title{A hands-on introduction to Physics-Informed Neural Networks for solving partial differential equations with benchmark tests taken from astrophysics
and plasma physics}
\author{Hubert Baty$$\footnote{Corresponding author: \href{mailto:hubert.baty@unistra.fr}{hubert.baty@unistra.fr}}~\\ {$$\small Observatoire Astronomique,
Université de Strasbourg, 67000 Strasbourg, France} \\ }
\date{\today}
\maketitle

\begin{abstract}

I provide an introduction to the application of deep learning and neural networks for solving partial differential equations (PDEs). The approach, known as physics-informed neural networks (PINNs), involves minimizing the residual of the equation evaluated at various points within the domain. Boundary conditions are incorporated either by introducing soft constraints with corresponding boundary data values in the minimization process or by strictly enforcing the solution with hard constraints. PINNs are tested on diverse PDEs extracted from two-dimensional physical/astrophysical problems. Specifically, we explore Grad-Shafranov-like equations that capture magnetohydrodynamic equilibria in magnetically dominated plasmas. Lane-Emden equations that model internal structure of stars in sef-gravitating hydrostatic equilibrium are also considered.
The flexibility of the method to handle various boundary conditions is illustrated through various examples, as well as its ease in solving parametric and inverse problems. The corresponding Python codes based on PyTorch/TensorFlow libraries are made available.

    \end{abstract}

\clearpage
\tableofcontents
\clearpage

\section{Introduction}

Since the introduction of Physics-Informed Neural Networks (PINNs) by Raissi et al. (2019), there has been a significant upsurge in interest in the PINNs technique, spanning various scientific fields. Notably, this technique offers several advantages, such as numerical simplicity compared to conventional schemes. Despite not excelling in terms of performance (accuracy and training computing time), PINNs present a compelling alternative for addressing challenges that prove difficult for traditional methods, such as inverse problems or parametric partial differential equations (PDEs). For comprehensive reviews, refer to Cuomo et al. (2022) and Karniadakis et al. (2021).

In this article, I introduce a tutorial on the PINNs technique applied to PDEs containing terms based on the Laplacian in two dimensions (2D), extending the previous tutorial work applied to ordinary differential
equations (ODEs) by Baty \& Baty (2023). More precisely, I focus on solving different Poisson-type equations called Grad-Shafranov equations and representing magnetic equilibria in magnetically dominated plasmas (e.g. the solar corona), following some examples presented in Baty \& Vigon (2024) and also other well known examples (see Cerfon et al. 2011).
Additionally, PDEs representative of two dimensional internal star structures are also solved, extending a previous work in a one dimensional approximation (Baty 2023a, Baty 2023b).
The latter equations are generally called
Lane-Emden equations in the literature. Note that, not only direct problems are considered but also inverse problems for which we seek to obtain an unknown term in the equation.
Of course, for these latter problems, additional data on the solutions is required.

I demonstrate how the PINNs technique is particularly well-suited when non Dirichlet-like conditions are imposed at boundaries. This is evident in scenarios involving mixed Dirichlet-Neumann conditions, especially those relevant to the Cauchy problem. 

The distinctive feature of the PINNs technique lies in minimizing the residual of the equation at a predefined set of data known as collocation points. At these points, the predicted solution is obligated to satisfy the differential equation. To achieve this, a physics-based loss function associated with the residual is formulated and employed. In the original method introduced by Raissi et al. (2019), often referred to as vanilla-PINNs in the literature, the initial and boundary conditions necessary for solving the equations are enforced through another set of data termed training points, where the solution is either known or assumed. These constraints are integrated by minimizing a second loss function, typically a measure of the error like the mean-squared error. This loss function captures the disparity between the predicted solution and the values imposed at the boundaries.
The integration of the two loss functions results in a total loss function, which is ultimately employed in a gradient descent algorithm. An advantageous aspect of PINNs is its minimal reliance on extensive training data, as only knowledge of the solution at the boundary is required for vanilla-PINNs. It's worth noting that, following the approach initially proposed by Lagaris (1998), there is another option to precisely enforce boundary conditions to eliminate the need for a training dataset (see for example Baty 2023b and Urban et al. 2023). This involves compelling the neural networks (NNs) to consistently assign the prescribed value at the boundary by utilizing a well-behaved trial function. The use of this latter second option, also referred as hard-PINNs below, is illustrated in this paper.

This paper is structured as follows. In Section 2, we begin by introducing the fundamentals of the PINNs approach for solving partial differential equations (PDEs). Section 3 focus on the application to solve
a simple Laplace equation in a rectangular domain, with the aim to compare the use of vanilla-PINNs versus hard-PINNs on problems involving Dirichlet BCs. The use of PINNs solvers on Poisson equations
with different types of BCs in rectangular domains are illustrated in Sect. 4. Section 5 focus on the application to a particular Helmholtz equation, representative of magnetic arcade equilibrium structures in
the solar corona. Another application for computing magnetic structures representative of curved loops is also considered as shown by the obtained solution of Grad-Shafranov equations in Section 6.
Section 7 is devoted to another astrophysical problem, that is solving Lane-Emden equations representative of the internal equilibrium structures of polytropic seff-gravitating gaz spheres (first order
approximation for the structure of stars). Finally, the use of PINNs for solving parametric differential equation and inverse problem are illustrated in Section 8 and Section 9 respectively, by
considering a stationary advection-diffusion problem. Conclusions are drawn in Section 10.

\section{Physics-Informed Neural Networks}\label{sec:PINNs}

\subsection{Problem statement for 2D direct problems}

We consider a partial differential equation (PDE) written in the following residual form as,
\begin{equation}
    \mathcal{F} (u, x, y, u_x, u_y, ...) = 0,\quad (x, y) \in \Omega, 
    \label{eq:PDE}
\end{equation}
where $u (x, y)$ denotes the desired solution and $u_x$, $u_y$, ... are the required associated partial derivatives of different orders with respect to $x$ and $y$.
Specific conditions must be also imposed at the domain boundary $\partial \Omega$ depending on the problem (see below in the paper).

Note that the $(x, y)$ space variables can also include non-cartesian coordinates (see Lane-Emden equation).

\subsection{Problem statement for parametric and inverse problems}

We consider a partial differential equation (PDE) written in the following residual form as,
\begin{equation}
    \mathcal{F} (u, x, \mu, u_x, ...) = 0,\quad x \in \Omega, \quad  \mu \in \Omega_p ,
    \label{eq:PDE}
\end{equation}
where the desired solution is now $u(x, \mu)$, with $x$ being the space variable associated to the one dimensional domain $\Omega$
and $\mu$ is a scalar parameter taking different values in $\Omega_p$. For parametric problems, $\mu$ is treated exactly as a second variable
in a 2D direct problem, but for inverse problems $\mu$ is consequently considered as an unknown. Boundary conditions (BC) are again necessary
for parametric problems, but additional conditions, such as knowledge of the solution at some $x$ values must be added for inverse problems.

Note that, for the sake of simplicity, we have considered only one dimensional space variable in this work for parametric and inverse problems.
The extension to higher spatial dimensions is however straightforward.

\subsection{Classical deep learning approach with neural networks using training data}

\begin{figure}[h]
    \centering
    \includegraphics[width=0.9\textwidth]{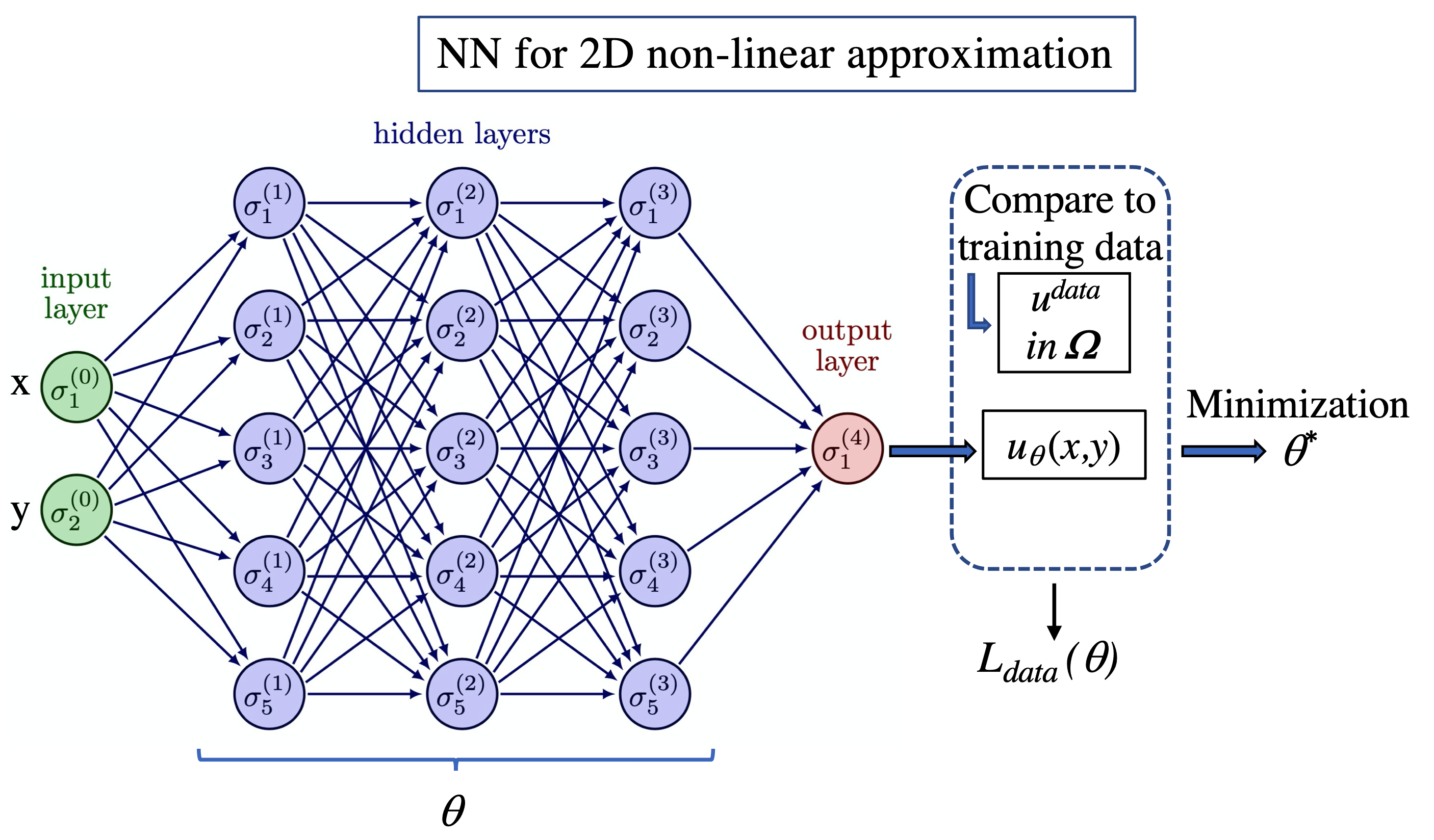}
    \caption{Schematic representation of the structure for a Neural Network (NN) applied to a non-linear approximation problem.
  The input layer has two input variables (i.e. two neurons)  for the two space coordinate variables $x$ and $y$.
    Three hidden layers with five neurons per layer are connected with the input and the output layer, where the latter has a single variable (one neuron) representing the predicted solution
    $u_\theta (x, y)$. The minimization procedure using the loss function $ L_{data}(\theta)$ is obtained by comparing $u_\theta$ to a training data set of values $u^{data}$ taken
    in the 2D domain $\Omega$. In this simplified example, $\theta$ represents a total number of $81$ scalar parameters.}
\end{figure}

In the classical deep learning approach with neural networks (NNs), the model is trained exclusively using available training data. This method involves feeding input data into the neural network, which then adjusts its internal parameters through a training process to minimize the difference between predicted and actual output values. The model learns patterns and relationships within the training dataset to make predictions on new, unseen data. This approach is common in various machine learning applications, where the emphasis is on leveraging labeled training examples to achieve accurate predictions. In this way, NNs serve as non-linear approximators.
 
\paragraph{Approximating the solution with a neural network.}
For any input $\bm{x}$ representing either the spatial coordinates $(x, y)$, or the combination of variables $(x, \mu$), or only $x$, depending on the problems, we want to be able to compute an approximation of the solution value $u(\bm{x})$ and eventually the parameter value $\mu$ (for inverse problems).

For this, we introduce what is called a multi-layer perceptron, which is one of the most common kind of neural networks. Note that any other statistical model could alternatively be used.
The goal is to calibrate its parameters $\theta$ such that $u_\theta$ approximates the target solution $u(\bm{x}) $.
$u_\theta$ is a non-linear approximation function, organized into a sequence of $L+1$ layers.
The first layer $\mathcal{N}^0$ is called the input layer and is simply:
\begin{equation}
    \mathcal{N}^0 (\bm{x}) = \bm{x}.
\end{equation}
Each subsequent layer $\ell$ is parameterized by its weight matrix $\bm{W}^\ell\in\mathbb{R}^{d_{\ell-1}\times d_\ell}$ and a bias vector $\bm{b}^\ell\in\mathbb{R}^{d_\ell}$, with $d_\ell$ defined as the output size of layer $\ell$.
Layers $\ell$ with $\ell\in \llbracket 1, L-1\rrbracket$ are called hidden layers, and their output value can be defined recursively:
\begin{equation}
    \mathcal{N}^\ell (\bm{x}) = \sigma (\bm{W}^\ell  \mathcal{N}^{\ell-1} (\bm{x}) +  \bm{b}^\ell),
\end{equation}
$\sigma$ is a non-linear function, generally called activation function.
While the most commonly used one is the ReLU ($\text{ReLU}(\bm{x}) = \max(\bm{x}, 0)$), we use the hyperbolic tangent $\tanh$ in this work, which is more suited than ReLU for building PINNs.
The final layer is the output layer, defined as follows:
\begin{equation}
    \mathcal{N}^L ( \bm{x}) = \bm{W}^L \mathcal{N}^{L-1} ( \bm{x}) +  \bm{b}^L ,
\end{equation}
Finally, the full neural network $u_\theta$ is defined as $u_\theta (\bm{x}) = \mathcal{N}^L (\bm{x})$.
It can be also written as a sequence of non-linear functions
\begin{equation}
u_\theta (\bm{x}) =  ( \mathcal{N}^L \circ \mathcal{N}^{L-1} ...\  \mathcal{N}^0) (\bm{x}),
\end{equation} where $\circ$ denotes the function composition and $\theta = \lbrace \bm{W}^l, \bm{b}^l \rbrace_{l=1,L}$ represents the parameters of the network.

\paragraph{Supervised learning approach using training data.}
The classical supervised learning approach assumes that we have at our disposal a dataset of $N_{data}$ known input-output pairs $(\bm{x}, u)$:
$$\mathcal{D} = \{(\bm{x}^{data}_i , u^{data}_i)\}_{i=1}^{N_{data}},$$
for $i\in\llbracket 1, N_{data}\rrbracket$.
$u_\theta$ is considered to be a good approximation of $u$ if predictions $u_\theta (\bm{x_i})$ are close to target outputs $u^{data}_i$ for every data samples $i$.
We want to minimize the prediction error on the dataset, hence it's natural to search for a value $\theta^\star$ solution of the following optimization problem:
\begin{equation}
    \theta^\star = \argmin_\theta L_{data}(\theta)
    \label{eq:learning-pb}
\end{equation}
with
\begin{equation}
    L_{data}(\theta) = \frac  {1} {N_{data} } \sum_{i=1}^{N_{data} } \left| ( u_\theta (\bm{x_i})  - u^{data}_i  \right|^2 .
\end{equation}
$L_{data}$ is called the loss function, and equation (7) the learning problem. It's important to note that the defined loss function relies on the mean
squared error formulation, but it's worth mentioning that alternative formulations are also possible. 
Solving equation (7) is typically accomplished through a (stochastic) gradient descent algorithm. This algorithm depends on automatic differentiation techniques to compute the gradient of the loss $L_{data}$
 with respect to the network parameters $\theta$. The algorithm is iteratively applied until convergence towards the minimum is achieved, either based on a predefined accuracy criterion or a specified maximum iteration number as,
\begin{equation}
    \theta^{j+1} =  \theta^{j} - l_r  \nabla_{\theta}  L (\theta^j) ,
\end{equation}
with $L = L_{data}$,
for the $j$-th iteration also called epoch in the literature, where $l_r$ is called the learning rate parameter.
In this work, we choose the well known Adam optimizer.
This algorithm likely involves updating the network parameters ($\theta$) iteratively in the opposite direction of the gradient to reduce the loss.
The standard automatic differentiation technique
is necessary to compute derivatives with respect to the
NN parameters, i.e. weights and biases (Baydin et al. 2018). This
technique consists of storing the various steps in the calculation
of a compound function, then calculating its gradient using the
chaine rule.
In practice, the learning process is significantly streamlined by leveraging open-source software libraries such as TensorFlow or PyTorch, especially when working with Python. These libraries provide pre-implemented functions and tools for building, training, and optimizing neural network models. TensorFlow and PyTorch offer user-friendly interfaces, extensive documentation, and a wealth of community support, making them popular choices for researchers and practitioners in the field of deep learning. Note that, in this work TensorFlow library is used for direct problems while PyTorch is preferred for parametric and inverse problems.

\begin{figure}[!h]
    \centering
    \includegraphics[width=\textwidth]{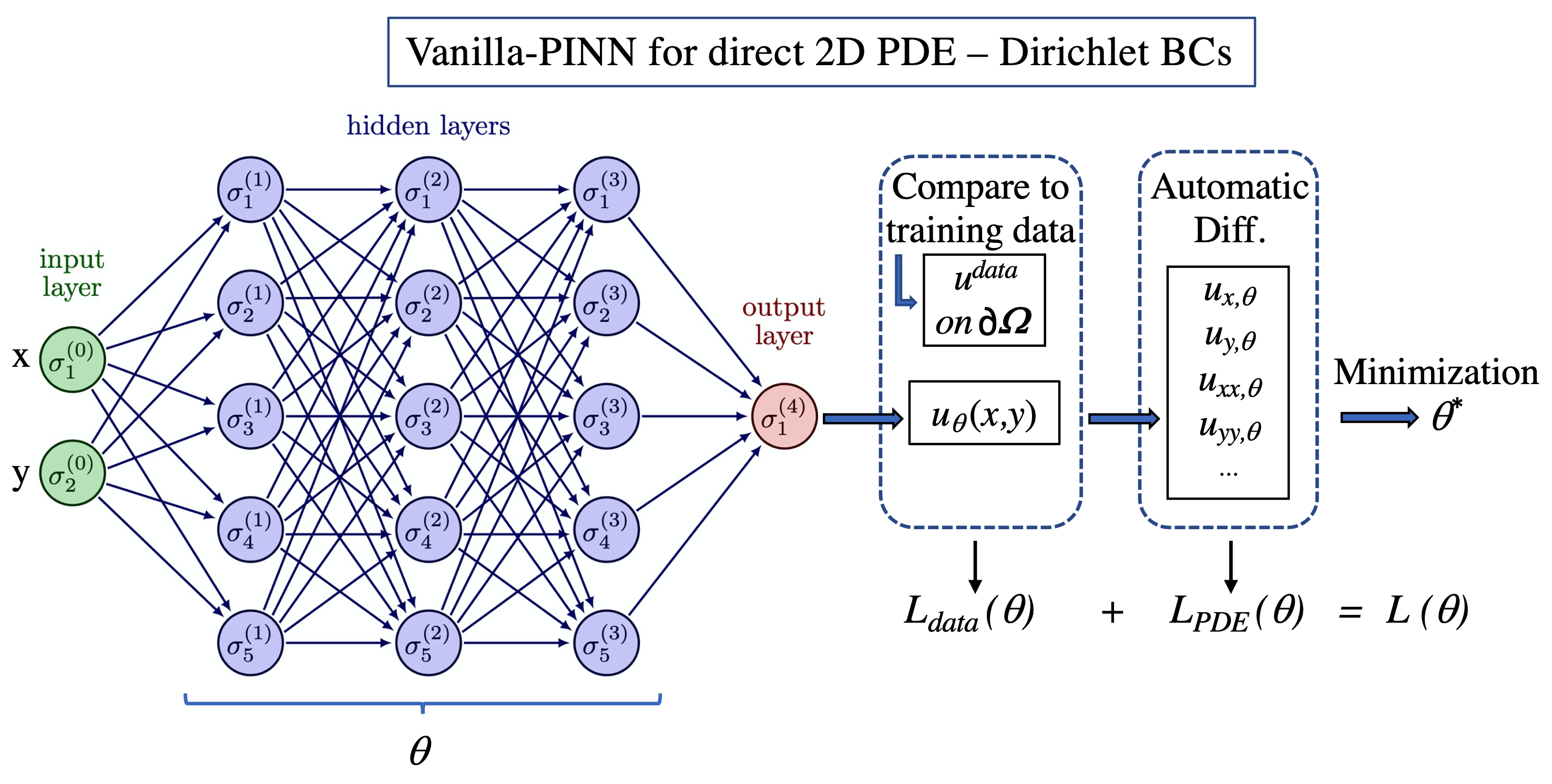}
    \caption{Schematic representation of the structure for a Physics-Informed Neural Network applied for solving a PDE associated to a 2D direct problem with Dirichlet-like BCs (soft constraints).
  The input layer has two input variables (i.e. two neurons)  for the two space coordinate variables $x$ and $y$.
    Three hidden layers with five neurons per layer are connected with the input and the output layer, where the latter has a single variable (one neuron) representing the predicted solution
    $u_\theta (x, y)$.  Automatic Differentiation (AD) is used in the procedure in order to evaluate the partial derivatives (i.e. $u_{x, \theta}$,  $u_{xy \theta}$...) necessary to form the PDE loss function $L_{ PDE} (\theta)$.
    The loss function $ L_{data}(\theta)$ is obtained with soft constraints (i.e. via the training data set) imposed on the boundary domain $\partial \Omega$.
     }
\end{figure}

A schematic representation of the architecture of a neural network designed to approximate a learned function $u (x, y)$ with a supervised learning approach using a training data set $N_{data}$
is visible in Fig. 1. In this case, two input neurons (first layer) represent the two spatial variables, and the output neuron (last layer) is the predicted solution $u_\theta$. 
The intermediate (hidden) layers between the input and output layers where the neural network learns complex patterns and representations, consists of $5$ neurons per layer in this example
of architecture. The total number of learned parameters is given by the formula $(i \times h_1 + h_1 \times h_2 + h_2 \times h_3 + h_3 \times o + h_1 + h_2 + h_3 + 0)$ that is therefore equal
to $81$, as $h_1 = h_2 = h_3 = 5$, $i = 2$ and $o =1$ ($i$ and $o$ being the number of input and output neurons respectively, and $h_i$ being the number of neurons of the $i-th$ hidden layer).

\subsection{PINNs for solving PDEs}

\begin{figure}[h]
    \centering
    \includegraphics[width=0.92\textwidth]{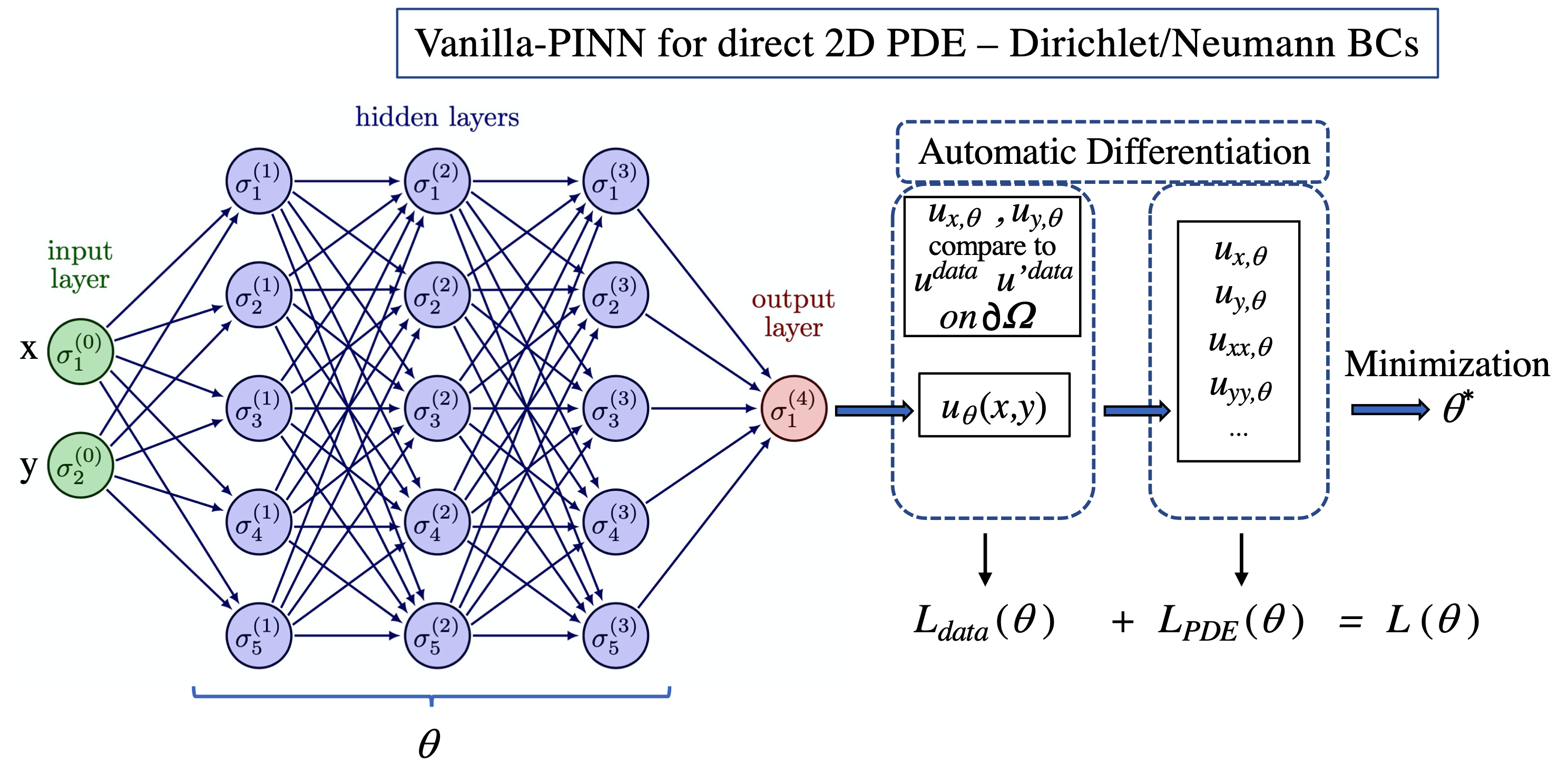}
    \caption{Schematic representation variant of previous figure where Neumann or Neumann-Dirichlet BCs are involved instead of purely Dirichlet BCs. The training data set includes
    additional knowledge on the exact derivatives $u^{'data}$.
    }
\end{figure}

\begin{figure}[h]
    \centering
    \includegraphics[width=0.92\textwidth]{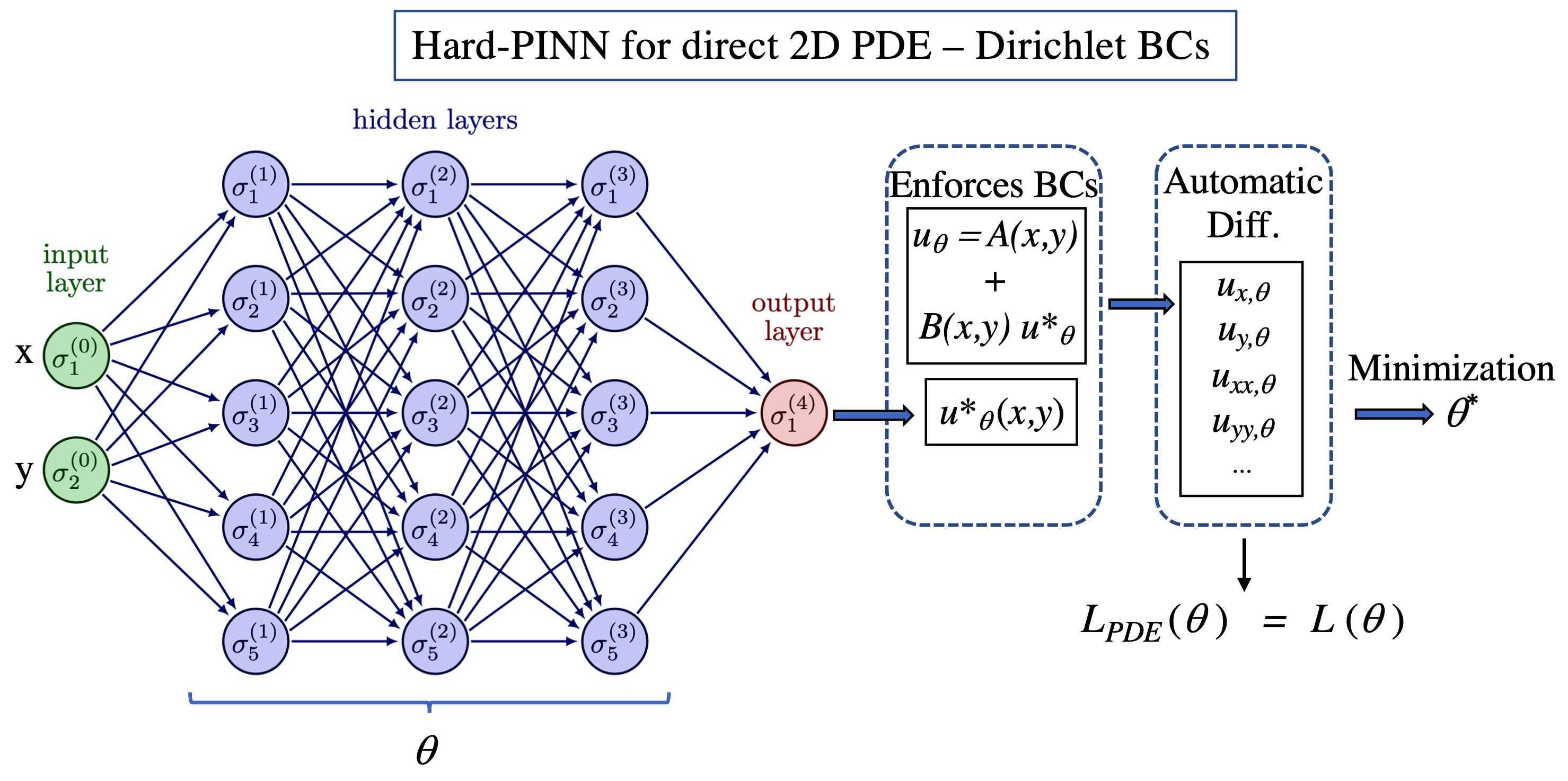}
    \caption{Schematic representation of a hard constraints BCs problem corresponding to previous figure (see text).  The boundary constraints are enforced via trial function. }
\end{figure}

\begin{figure}[h]
    \centering
    \includegraphics[width=0.92\textwidth]{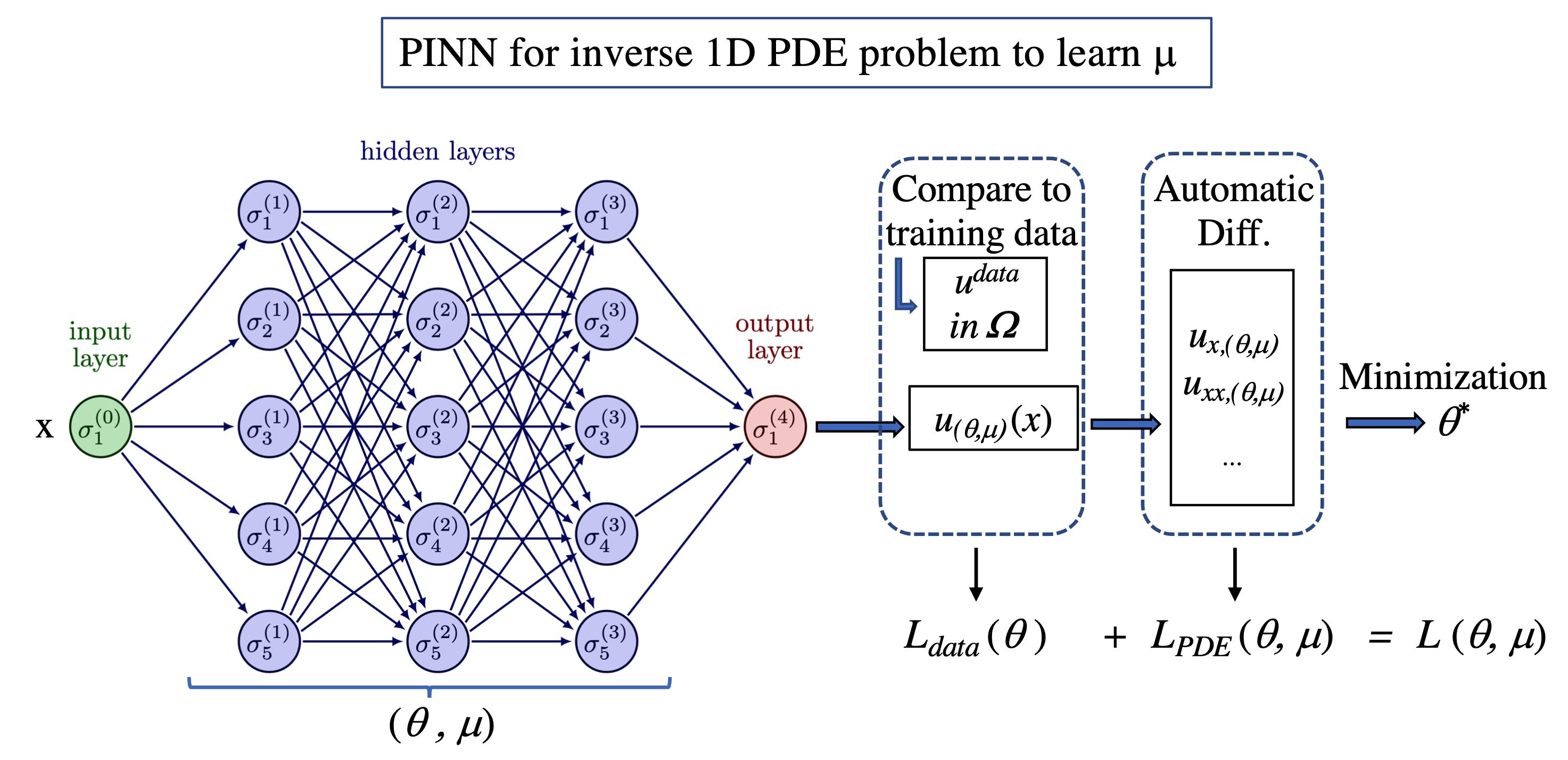}
    \caption{Schematic representation of an inverse problem. A 1D differential equation is considered (ODE in fact) parametrized with an unknown coefficient $\mu$
    to be discovered. The training data set must include data from the whole domain $ \Omega$ (not only at boundaries).}
\end{figure}

In PINNs approach, a specific loss function $L_{PDE}$ can be defined as,
\begin{equation}
    L_{ PDE} (\theta) = \frac  {1} {N_c} \sum_{i=1}^{N_c}  \left|  \mathcal{F} ( u_\theta (\bm{x_i} )  )   \right|^2 ,
\end{equation}
where the evaluation of the residual equation is performed on a set
of $N_c$ points denoted as $\bm{x}_i$. These points are commonly referred to as collocation points.
A composite total loss function is typically formulated as follows
\begin{equation}
    L(\theta) = \omega_{data} L_{data} (\theta) +  \omega_{PDE} L_{PDE} (\theta),
\end{equation}
where $\omega_{data}$ and  $\omega_{PDE}$ are weights to be assigned to ameliorate
potential imbalances between the two partial losses. These weights can be user-specified or automatically tuned.
In this way, the previously described gradient descent algorithm given by equation (9) is applied to iteratively reduce the total loss.
By including boundary data in the training dataset, the neural network can learn to approximate the solution not only
within the domain but also at the boundaries where the known solutions are available. PINNs are thus well-suited for solving
PDEs in inverse problems in a data-driven manner.

In the context of solving PDEs in direct and parametric problems, 
for cases involving purely Dirichlet BCs, the training data set is generally reduced
to the sole solution value at the boundary. However, for problems involving Neumann BCs or a combination of Neumann and Dirichlet BCs (mixed Neumann-Dirichlet)
the training dataset must extend beyond simple solution values. In these cases (Neumann BCs), since Neumann BCs involve the specification of perpendicular derivative values at the boundaries,
 the training dataset needs to include information about these derivatives. This ensures that the neural network learns to capture the behavior of the solution
 with respect to the normal derivative at the boundary. In other cases (Mixed Neumann-Dirichlet BCs), the training dataset should incorporate solution values at Dirichlet boundaries
 and also derivative information at Neumann boundaries.

A schematic representation of the architecture of a neural network designed to solve a PDE using PINNs is visible in Fig. 2. For simplicity, the weights are taken to be unity
in this example.This scheme corresponds to a direct problem involving Dirichlet BCs imposed with soft constraints, as the solution on the boundary is not exactly enforced in this way.
A similar problem involving Neumann BCs or mixed Neumann-Dirichlet BCs is schematized in Fig. 3. Note that, the procedure used to impose the derivative values
for Neumann BCs is slightly different from the one proposed in Baty (2023b). Indeed, in the latter work  the first collocation point was used (see Eq. 5 in the manuscript),
contrary to the present work where the derivative values are part of the training data set.

As explained in introduction, an alternative option to exactly enforce BCs is to employ a well-behaved trial function (Lagaris 1998), $u_\theta (x,y)= A(x,y) + B(x,y) u_\theta^* (x,y)$, where
now $u_\theta^* (x,y)$ is the output value resulting from the NN transformation to be distinguished from the final predicted solution $u_\theta (x,y)$.
In this way the use of a training data set is not
necessary, and only $L_{PDE} (\theta)$ survives to form the used loss function $L(\theta)$.
The latter variant using thus hard boundary constraints is schematized in Fig. 4.

 For a 1D spatial (i.e. $x$) parametric problem, $y$ must be replaced by $\mu$, and the problem is thus similar to the 2D spatial problems schematized in Figs 2-4.
 
 Finally, for a 1D inverse problem, only one input neuron is needed for the space coordinate ($x$), and the unknown parameter $\mu$ is learned exactly as an additional network
 parameter like the weights and biases $\theta$ (see further in the paper), as can be seen in the schematic representation of Fig. 5.

\subsection{Python codes}
The TensorFlow/Pytorch Python-based codes and data-sets accompanying this manuscript are available on the GitHub repository\footnote{\url{https://github.com/hubertbaty/PINNS-PDE}}.
We have chosen to use very simple deep feed-forward networks architectures with hyperbolic activation functions. In this work, the tuning of the architecture of the network
(number of hidden layers, number of neurons per layer) and of the other hyper-parameters (learning rate, loss weights) is done manually.
Although more systematic/automatic procedures could be used, this is a difficult task not considered in this work. Note that, for the examples considered in this work, the number
of employed hidden layers can typically vary between $4$ and $7$ and the number neurons per layer between $20$ and $40$. The chosen learning rate is also taken
in the range varying between $10^{-4}$ and $10^{-3}$, as a too small value leads to a very slow convergence and a too high value to convergence difficulties due to strong
oscillations in the gradient descent algorithm. The choice of the exact optimizer variant is also important, and the best (and simpler) option is generally provided by the Adam optimizer.

\section{Laplace equation}

In order to illustrate a first practical implementation of the method, we consider a simple Laplace equation in 2D cartesian coordinates:
\begin{equation}
    u_{ xx} +  u_{ yy} = 0 ,
\end{equation}
with $u_{ xx}$ and $u_{ yy}$ being the second order derivatives of $u$ with respect to $x$ and $y$ respectively. The integration domain
is a square with $\Omega = [-1, 1] \times [-1, 1]$. We also focus on a case having the exact solution $u (x, y) = x^2 - y^2$. Of course, the Dirichlet
BCs must match the correct exact solution values.

For Laplace equation, I propose to use the two variants of PINNs technique on a Dirichlet BCs problem, namely the vanilla-PINNs and hard-PINNs,
and comparing their results.

\subsection{Using vanilla-PINNs on Dirichlet BCs problem}

\begin{figure}[h]
    \centering
    \includegraphics[width=0.4\textwidth]{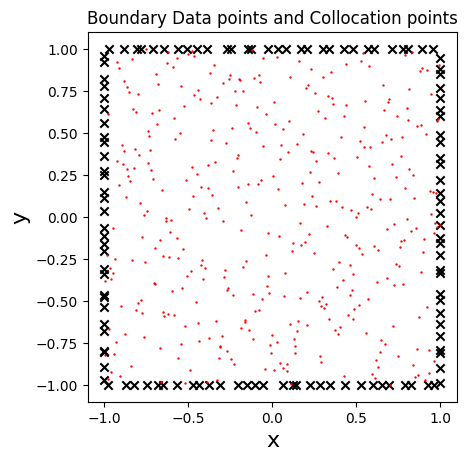}
    \includegraphics[width=0.48\textwidth]{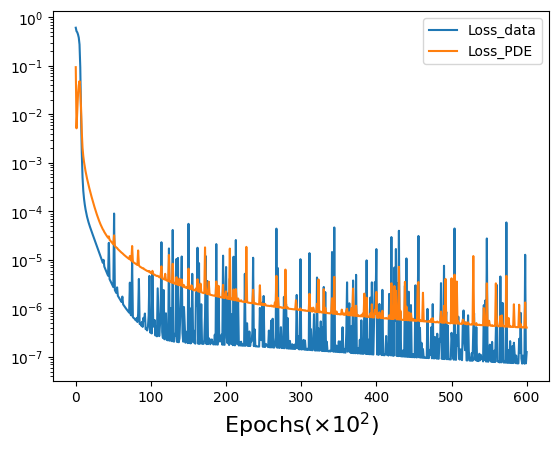}
    \caption{(Left panel) Distribution of data sets showing the space localization of training data points at the four boundaries (i.e. $N_{data} = 120$) and collocation points (i.e. $N_c = 400$) inside the domain,
    for solving Laplace problem using vanilla-PINNs. (Right panel) Evolution of the two partial losses
$L_{data}$ and $L_{PDE} $ as functions of the number of iterations (i.e. epochs).
    }
\end{figure}

\begin{figure}[h]
    \centering
    \includegraphics[width=\textwidth]{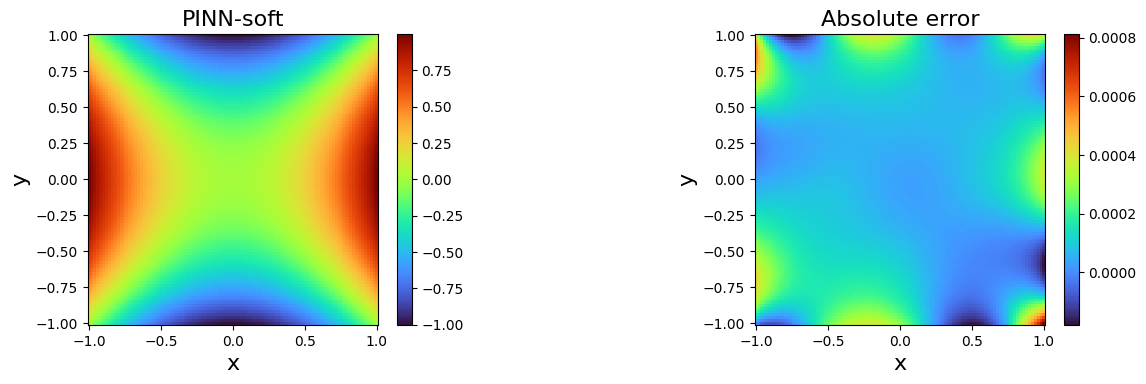}
  \caption{ Solution (left panel) and absolute error (right panel) distributions as colored iso-contours corresponding to problem associated to the previous figure.
      }
\end{figure}

\begin{figure}[h]
    \centering
    \includegraphics[width=\textwidth]{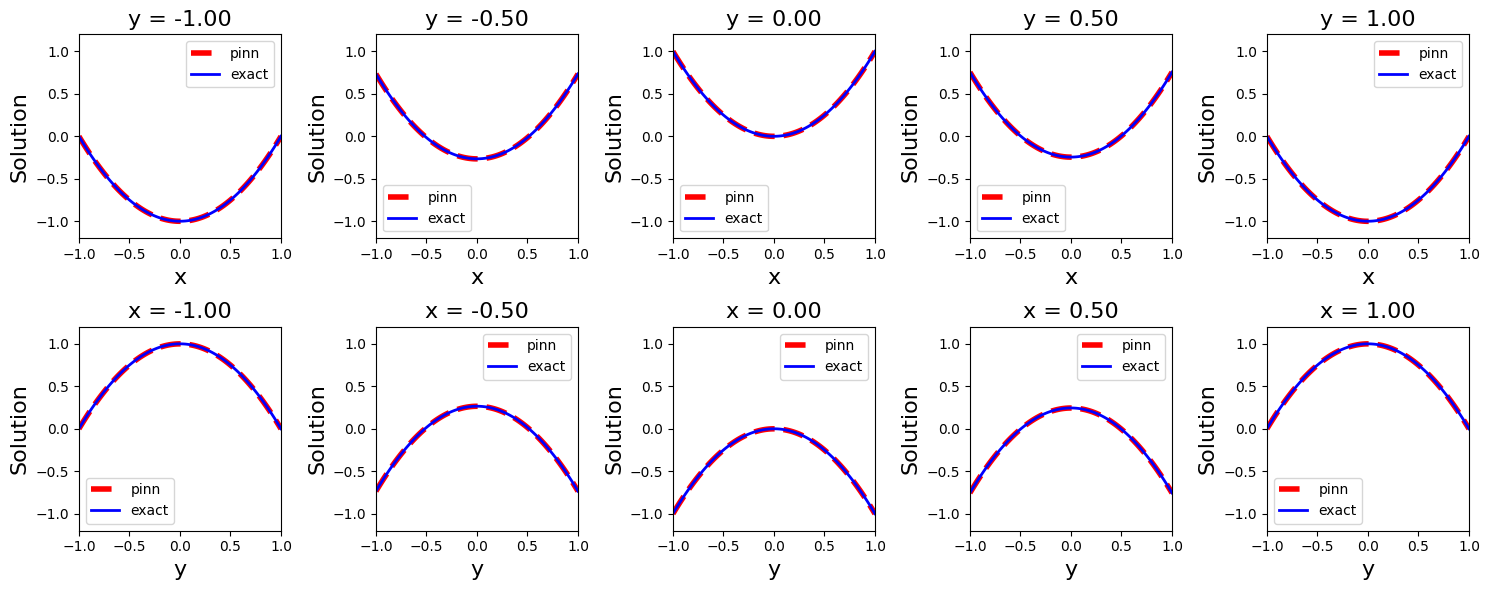}
  \caption{One-dimensional solution obtained for different $x$ and $y $ particular values compared with the exact analytical solution (plain line) corresponding
  to the previous figure.
    }
\end{figure}

\begin{figure}[h]
    \centering
    \includegraphics[width=0.36\textwidth]{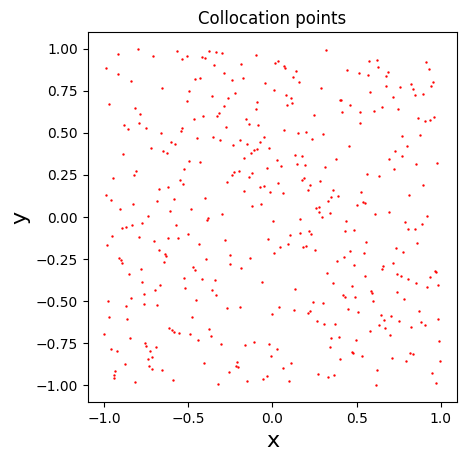}
    \includegraphics[width=0.4\textwidth]{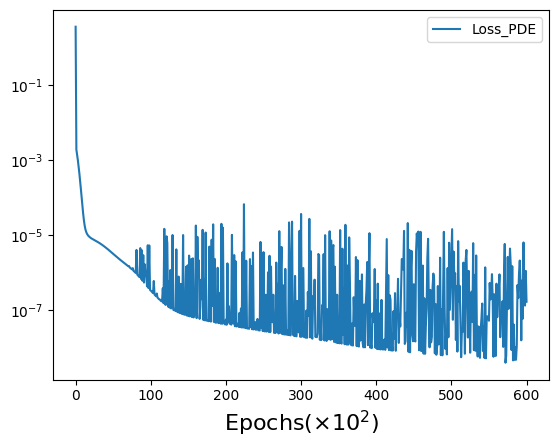}
    \caption{(Left panel) Distribution of data sets showing collocation points (i.e. $N_c = 400$) inside the domain,
    for solving Laplace problem using hard-PINNs. (Right panel) Evolution of the total loss $L = L_{PDE} $ as function of the number of iterations (i.e. epochs).
    }
\end{figure}

\begin{figure}[h]
    \centering
    \includegraphics[width=0.8\textwidth]{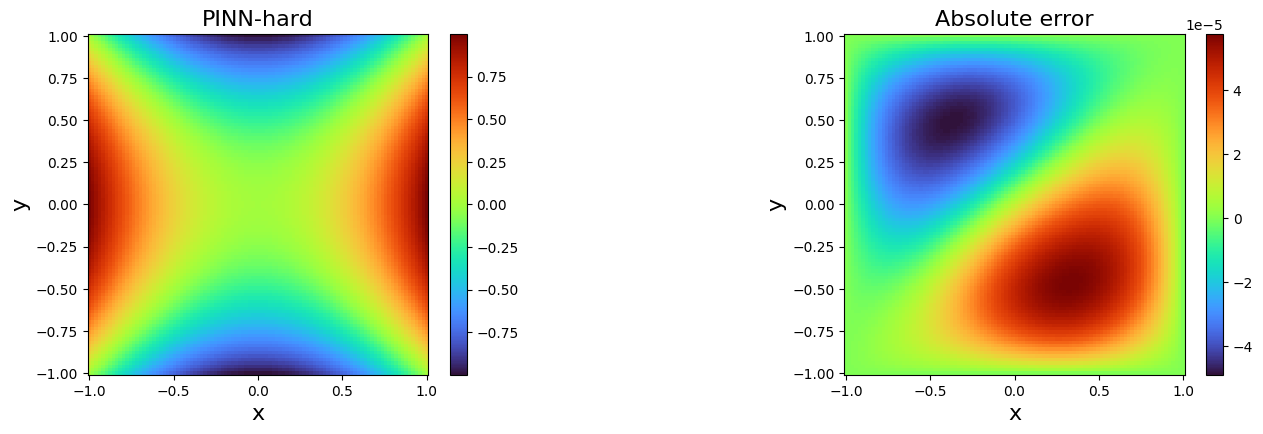}
  \caption{ Solution (left panel) and absolute error (right panel) distributions as colored iso-contours corresponding to problem associated to the previous figure.
      }
\end{figure}

First, we need to generate a data set of training data with points localized at the $4$ boundaries, i.e. $x \pm 1$ and $y \pm 1$.
In this example we choose $120$ points ($30$ per boundary) with a randomly distribution. We also need to generate the set of
collocation points inside the domain, e.g. $400$ randomly distributed points are used. Note that, I use a pseudo-random distribution
(Latin-hypercube strategy) in order to avoid empty regions. The resulting data generation is illustrated in Fig. 6 (left panel).

We apply the PINN algorithm schematized in Fig. 2. 
The evolution of the two loss functions with the training epochs that are reported in Fig. 6 (right panel), shows the convergence toward the predicted solution,
as very small values are obtained at the end. Note that the training is stopped after $60000$ epochs. For this problem, I
 have chosen a network architecture having $5$ hidden layers with $20$ neurons per layer. This represents $1761$ learned parameters. A learning rate of
 $l_r = 2 \times 10^{-4}$ is also chosen. The latter parameters choice slightly influences the results but is not fundamental as long as the number of layers/neurons is not too small (Baty 2023a). 
A faster convergence can be also obtained by taking a variable learning rate with a decreasing value with the advance of the training process. 

The solution and the error distribution at the end of the training are plotted in Fig. 7 exhibiting a maximum absolute error of order $0.0008$,
This is well confirmed by inspecting corresponding one dimensional cuts comparing predicted and exact solutions, as can be seen in Fig. 8.
 Note that the predicted PINNs solution and associated error distribution are obtained using a third set of points (different from the collocation points) that is taken to be a uniform grid of $100 \times 100$ points here, otherwise the error could be artificially small (overfitting effect). One must also note that the error is higher near the boundary due to the coexistence of data/collocation points in these regions. In this way, once trained, the network allows to predict the solution quasi-instantaneously at any point inside the integration domain, without the need for interpolation (as done e.g. with finite- difference methods when the point is situated between two grid points). The precision of PINNs is known to be very good but less than more traditional methods (e.g. like in finite-element codes). This is a general property of minimization techniques based on gradient descent algorithms (Press et al. 2007; Baty 2023). However, a finer tuning of the network parameters together with the introduction of optimal combinations for weights of the partial losses can generally ameliorate the results, which is beyond the scope of this work.
 
Traditional numerical schemes are generally characterized by some convergence order due to the space discretization. For example, a method of order two means that the associated
truncation error is divided by $4$ when the space discretization factor is divided by $2$, resulting in a precise given decreasing parabolic scaling law. PINNs are
statistical methods that are consequently not expected to follow such law. More precisely, multiplying by $2$ (for example) the number of collocation points and/or
the number of training data points does not necessarily lead to a dependence law for the error. The only well established result is that, there is a minimum number
of points (depending on the problem) necessary for convergence towards an acceptable solution. There is also a minimum number of hidden layers and number of
neurons per layer. A too large architecture can even degrade the precision of the results.
One can refer to tests done on Lane-Emden ODEs on the latter property (Baty 2023a, Baty 2023b).

\subsection{Using hard-PINNs on Dirichlet BCs problems}

We consider the same problem as in the previous sub-section, but following instead the variant schematized in Fig. 4. Only
the collocation data set is necessary, and generated in the same way as previously using vanilla-PINNs. However, the predicted solution
$u_\theta$ now differs from the function $u^*_{\theta}$ resulting from the output neural network transformation. Indeed, following Lagaris (1998),
the predicted solution $u_\theta$  can be written as,
\begin{equation}
    u_\theta = A (x,y) + B (x,y) u^*_{\theta} , 
\end{equation}
where the function $A (x,y)$ is designed to exactly satisfy the BCs without any adjustable parameter, and the remaining form $B (x,y) u^*_{\theta}$
is constructed so as to not contribute to the BCs. The ajustable parameters are thus contained in the neural network output function only that is now
$u^*_{\theta}$, and the function $B(x,y)$ must vanish at the boundaries. The choice of the two functions $A$ and $B$ is not unique and can affect
the efficiency of the algorithm.

In the present example, following prescriptions given by Lagaris (1998),
we have $B = (1 -x ) (1 + x) (1 - y) (1 + y) $ and $A = (1 - y^2) + (x^2 - 1)$. This is $u_\theta$ that is used to evaluate
the loss function $L(\theta) = L_{PDE} (\theta)$ in the minimization procedure, as it should satisfy the PDE contrary to $u^*_{\theta}$.
The loss function is based on the residual that is simply $  \mathcal{F} =   u_{ xx} +  u_{ yy}$.
Note that, it is not always simple (or even possible) to define the two functions $A$ and $B$, especially when the boundaries are more complex or/and the BCs
involve Neumann conditions (see Lagaris 1998).

Following the PINN algorithm schematized in Fig. 4, we can thus solve the same previous Laplace equation.
The results are plotted in Fig. 9 and Fig. 10. Note that only collocation points are now necessary.
The precision is better compared to results obtained using vanilla-PINNs, as the maximum absolute error is now of order $5 \times 10^{-5}$.
Moreover, the error is mainly localized inside the domain. This was expected as the boundary values are exactly imposed with this variant.
However, as explained in the previous figure, this is not always easy/possible to use this method, contrary to the vanilla-PINN variant.

\section{Poisson equations}

In this section, we test the method on Poisson-type equations in 2D cartesian coordinates. Thus, we consider the following form,
\begin{equation}
    u_{ xx} +  u_{ yy} = f(x, y) ,
\end{equation}
having the following $5$ manufactured exact solutions taken from Nishikawa (2023):
\begin{equation}
\begin{split}
&  (a)   \  \   \  \   \  \  u (x, y) =     e^{ xy}              \\
&  (b)   \  \   \  \   \  \  u (x, y) =       e^{kx} \sin(ky)  + \frac{1} {4}   (x^2 + y^2)            \\
& (c)   \  \   \  \   \  \  u (x, y) =     \sinh(x)               \\
&  (d)   \  \   \  \   \  \  u (x, y) =    e^{ x^2 + y^2}                \\
&  (e)  \  \   \  \   \  \  u (x, y) =   e^{ xy} +  \sinh(x)  
 \end{split}
\end{equation}
for respectively,
\begin{equation}
\begin{split}
&  (a)   \  \   \  \   \  \  f (x, y) =     e^{ xy}    (x^2+y^2)          \\
&  (b)   \  \   \  \   \  \  f (x, y) =       1            \\
& (c)   \  \   \  \   \  \  f (x, y) =     \sinh(x)               \\
&  (d)   \  \   \  \   \  \  f (x, y) =    4(x^2+y^2 + 1) e^{ x^2 + y^2}                \\
&  (e)  \  \   \  \   \  \  f (x, y) =   e^{ xy}    (x^2+y^2)  +  \sinh(x)  
 \end{split}
\end{equation}
We consider the integration domain $\Omega = [0, 1] \times [0, 1]$ with different types of boundary conditions specified below.

In this section, I propose to compare the use of different BCs with vanilla-PINNs variant (only) on Poisson problems. Now, the
loss function for PDE is based on the residual equation $ \mathcal{F} =   u_{ xx} +  u_{ yy} - f(x, y)$.

\subsection{Using vanilla-PINNs on Dirichlet problems}

\begin{figure}[h]
    \centering
    \includegraphics[width=0.8\textwidth]{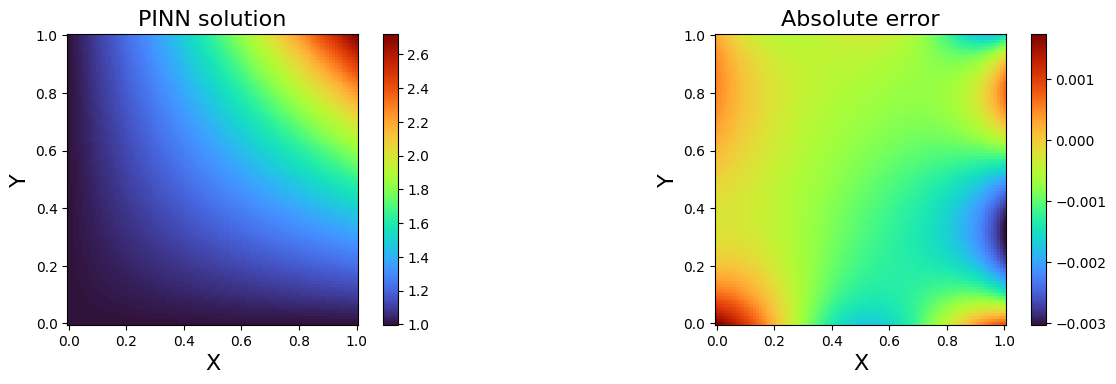}
  \caption{ Vanilla-PINNs solution (left panel) and associated absolute error (right panel) distributions as colored iso-contours corresponding to Poisson-problem (case-a) equation
  with Dirichlet BCs.
      }
\end{figure}

\begin{figure}[h]
    \centering
    \includegraphics[width=0.4\textwidth]{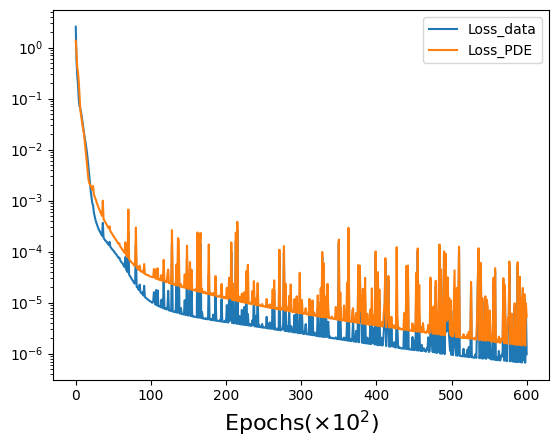}
    \caption{Evolution of the two partial losses $L_{data}$ and $L_{PDE} $ as functions of the number of iterations (i.e. epochs) for Poisson equation (case-a)
    with Dirichlet BCs using vanilla-PINN, associated to previous figure.
    }
\end{figure}

First, we need to generate a data set of training data with points localized at the $4$ boundaries, i.e. $x=0$, $x = 1$,  $y = 0$, and $y = 1$.
As illustrated for Poisson problem, we choose $120$ points ($30$ per boundary) with a randomly distribution. We also need to generate the set of
collocation points inside the domain, e.g. $400$ randomly distributed points (with Latin-hypercube strategy) are used. 

We apply the PINN algorithm schematized in Fig. 2, to solve the equation for the first case (a).
The evolution of the two loss functions with the training epochs that are reported in Fig. 11, shows the convergence toward the predicted solution,
as very small values are obtained at the end. Note that the training is stopped after $60000$ epochs. 
For such Poisson PDE the PDE loss function is taken to be based on $\mathcal{F} =   u_{ xx} +  u_{ yy} - f(x, y)$.
For this problem, I
 have chosen a network architecture having $6$ hidden layers with $20$ neurons per layer. This represents $2181$ learned parameters. A learning rate of
 $l_r = 3 \times 10^{-4}$ is also chosen.

The solution and the error distribution at the end of the training are plotted in Fig. 12 exhibiting a maximum absolute error of order $0.003$
(the corresponding maximum relative error is similar to the previous Poisson problem). Similar results are obtain for the other equations, as one can see on
results reported in appendix.

Of course, if the BCs are applied only on a part of the whole boundary $\partial \Omega$ (for example on $3$ or even on $2$ of the four boundaries),
the solution can be also obtained but with a lower accuracy.

\subsection{Using vanilla-PINNs on Neumann problems}

Now, we solve the same Poisson equation using Neumann BCs at the $4$ boundaries. The results show a convergence towards a wrong solution
that is the expected exact solution up to an additive constant (the latter value being $3.47$ for case-a, see Figs. 13-14).

This is not surprising as the solution to Poisson problem subject to pure Neumann conditions is known to be unique only
up to an overall additive constant. One can check this is also the case solution to other Poisson equations (cases b-e), as plotted
in Appendix.

This example illustrates that, the convergence of the loss function towards a low value does not guarantee obtaining the exact solution.

\begin{figure}[h]
    \centering
    \includegraphics[width=0.8\textwidth]{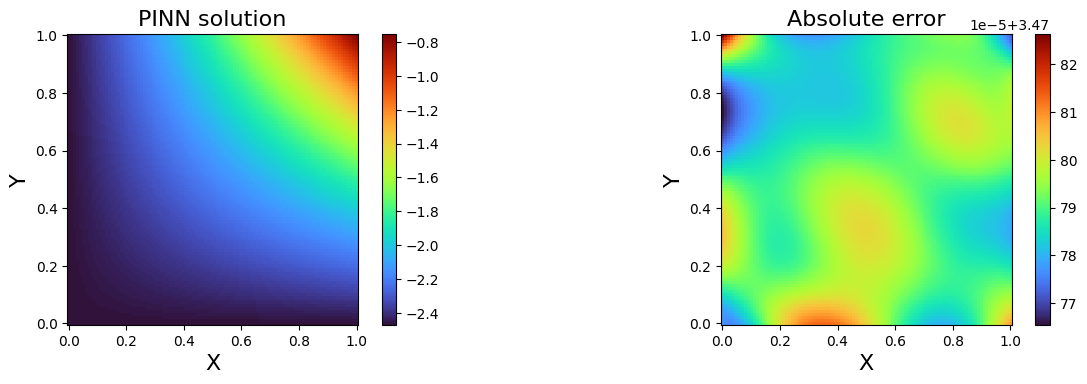}
  \caption{ Solution (left panel) and absolute error (right panel) distributions as colored iso-contours corresponding to the Poisson problem (case-a) with
  Neumann BCs and vanilla-PINN.
      }
\end{figure}

\begin{figure}[h]
    \centering
    \includegraphics[width=0.4\textwidth]{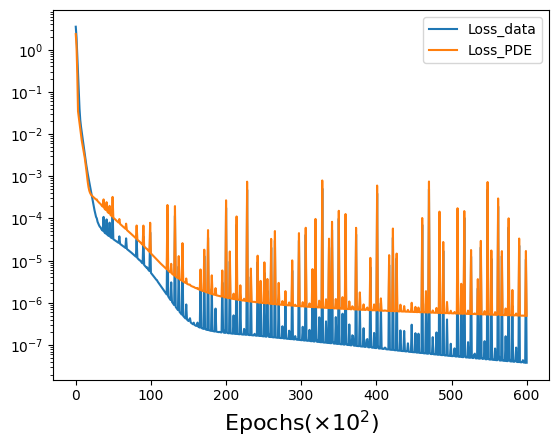}
    \caption{Evolution of the two partial losses $L_{data}$ and $L_{PDE} $ as functions of the number of iterations (i.e. epochs) for Poisson equation (case a)
    with Neumann BCs using vanilla-PINN.
    }
\end{figure}

\subsection{Using vanilla-PINNs on Neumann-Dirichlet problems}

Using mixed boundary conditions (i.e. Dirichlet on part on boundary $\partial \Omega$ and Neumann on the rest) lead to the exact solution
as for purely Dirichlet BCs. An example of results obtained for the first Poisson equation (a-case), where Dirichlet BC values are imposed at $x = 1$ and $y=1$ and
Neumann BC values at the $2$ other boundaries ($x=0$ and $y=0$), is illustrated in Fig. 15.

Although it depends oc cases considered, the error value is similar with (generally) a slightly better accuracy when compared to results for purely Dirichlet BCs.

\begin{figure}[h]
    \centering
    \includegraphics[width=0.8\textwidth]{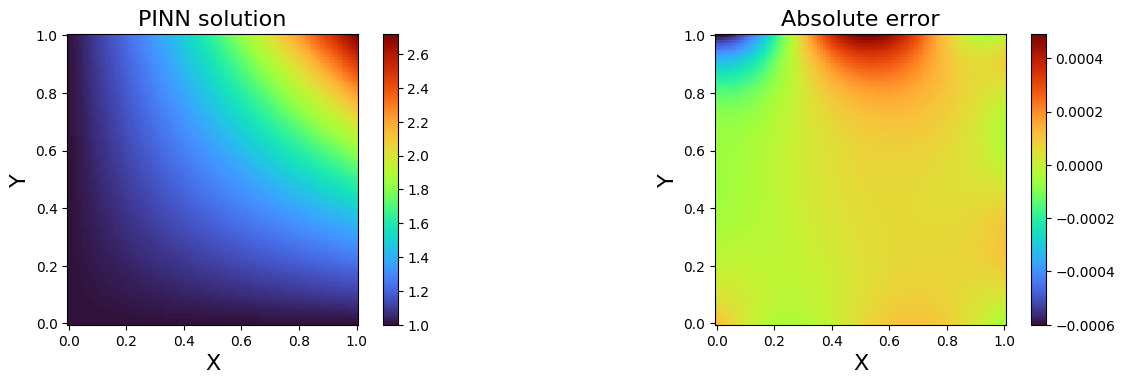}
    \caption{ Solution (left panel) and absolute error (right panel) distributions as colored iso-contours corresponding to the Poisson problem (case a) with
  mixed Dirichlet-Neumann BCs and vanilla-PINN (see text).
      }
\end{figure}

\subsection{Using vanilla-PINNs on Cauchy problems}

We consider Cauchy boundary conditions (where both the solution and the perpendicular derivative are simultaneously specified on
the same boundary). For example, for the results obtained in Fig. 16 on the case (a), we have imposed the exact solutions and
perpendicular derivatives at the two boundaries $x = 0$ and $x = 1$ only. The accuracy appears to be similar to solution obtained for the same problem 
using mixed boundary conditions (see previous sub-section).

\begin{figure}[h]
    \centering
    \includegraphics[width=0.8\textwidth]{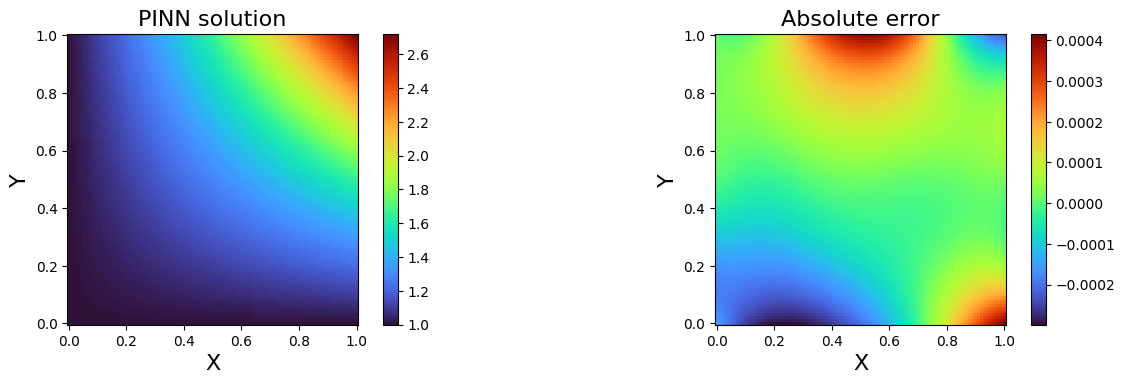}
    \caption{ Solution (left panel) and absolute error (right panel) distributions as colored iso-contours corresponding to the Poisson problem (case a) with
  Cauchy BCs and vanilla-PINN (see text).
      }
\end{figure}

\section{Helmholtz equations}

\subsection{Specific problem related to astrophysics}

\begin{figure}[h]
    \centering
    \includegraphics[width=0.45\textwidth]{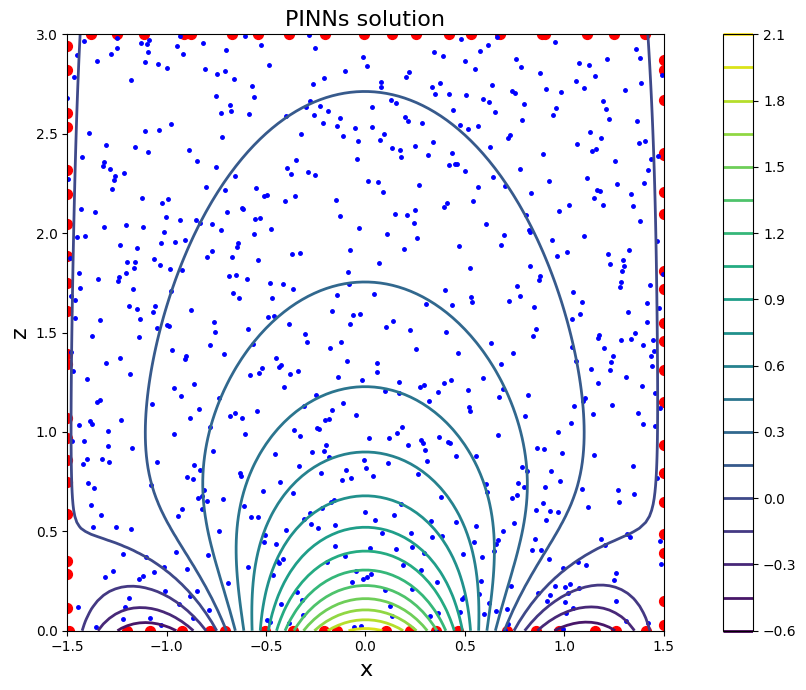}
     \includegraphics[width=0.45\textwidth]{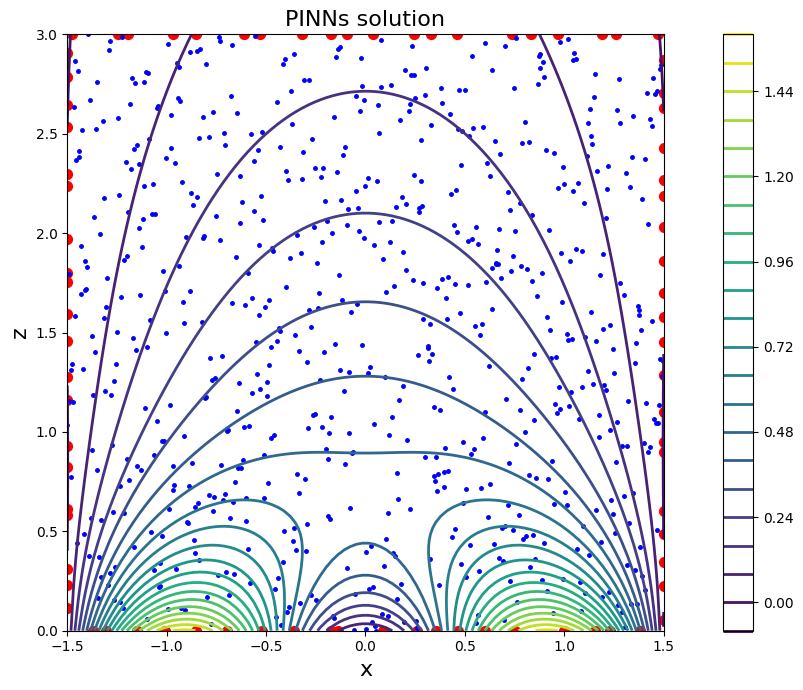}
    \caption{ Solutions (iso-contours of $ u (x, z)$) predicted by PINNs solver for the Helmholtz equation for two cases, i.e. for the parameters combination $(a_1, a_2, a_3)$ 
    equal to  $(1, 0, 1)$ and $(1, 0 , -1)$ in left and right panel respectively. The distribution of data sets (training and collocation points) is also indicated using
    red and blue dots at the boundary and inside the domain respectively.
      }
\end{figure}

\begin{figure}[h]
    \centering
    \includegraphics[width=0.45\textwidth]{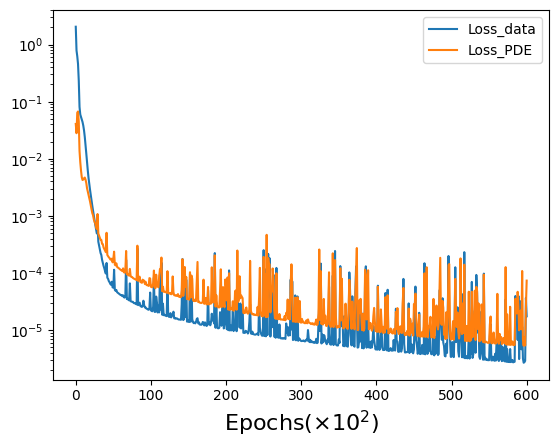}
    \includegraphics[width=0.45\textwidth]{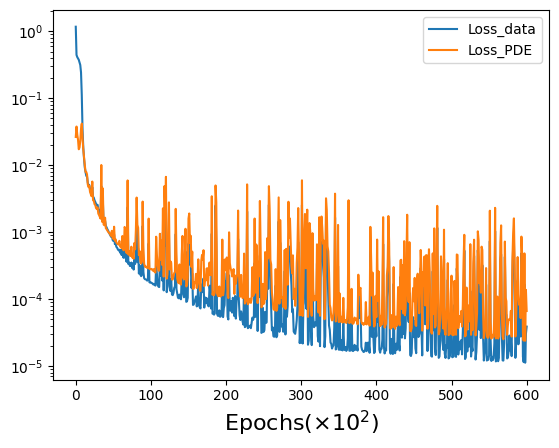}
    \caption{ Evolution of the losses (training data and PDE) during the training as functions of epochs, corresponding to the two cases respectively (see previous figure).
          }
\end{figure}

\begin{figure}[h]
    \centering
    \includegraphics[width=0.99\textwidth]{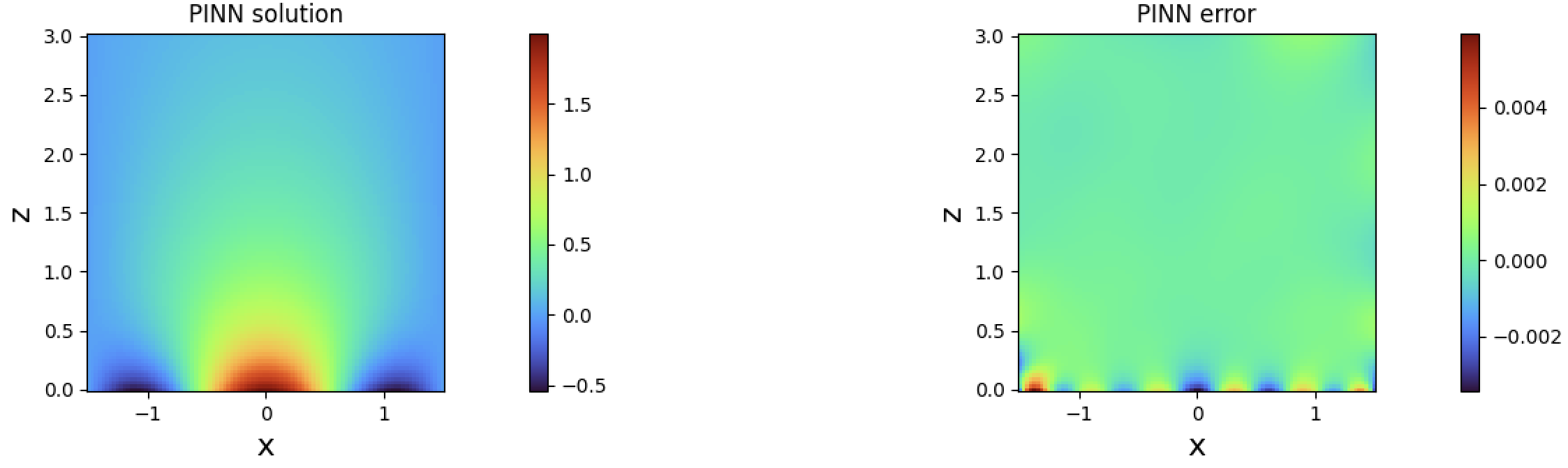}
    \caption{Predicted solution (left panel) and absolute error distribution (right panel)  using colored iso-contours, corresponding to the first case 
    parameters combination $(a_1, a_2, a_3)$ equal to  $(1, 0, 1)$ of the Helmholtz problem (see text and the
    two previous figures).
      }
\end{figure}

Following a previous work (Baty \& Vigon 2024), we solve an Helmholtz equation,
    \begin{equation}
       \Delta  u + c^2  u = 0,
      \end{equation}
where $c$ is a constant and $ \Delta =  \frac {\partial^2} {\partial x^2} +  \frac {\partial^2} {\partial z^2}$ is the cartesian Laplacian operator.
The spatial integration domain with $(x, z)$  extending between $[- \frac {L} {2}, \frac {L} {2}]$ respectively is taken.

In an astrophysical context, this equation represents an equilibrium magnetic structure. This is for example the case of
magnetic arcades observed in the solar corona. The scalar function $u (x,z)$ is known as a flux function allowing to deduce
the magnetic field vector component in the $(x, z)$ plane and magnetic field lines, via isocontour values of $u$.
The sun surface is situated at $z = 0$ as $z$ represents the altitude. 
The remaining magnetic field component $B_y$ being simply added to
the previous one in case of translational symmetry generally assumed in this invariant direction. On other words, the total magnetic
field may be written as,
    \begin{equation}
      {\bf B }  (x, z) = \nabla u (x, z) \times {\bf e_y } + B_y (u)  {\bf e_y } ,
      \end{equation}
where $ {\bf e_y }$ is the unit vector of the cartesian basis along the $y$ direction. 
Exact solutions for triple arcade structures can be obtained using Fourier-series as
      \begin{equation}
         u (x, z) =   \sum_{k=1}^{3}  \exp ( -  \nu z)   \left [   a_k \cos  \left ( \frac {k \pi}  {L} x  \right )  \right ] .
      \end{equation}
The latter solution is periodic in $x$, and
the relationship $\nu^2 =  \frac {k^2 \pi^2}  {L^2} - c^2$ applies as a consequence of the above Helmholtz equation. More details about the context
can be found in Baty  \& Vigon (2024) and references therein.

\subsection{Solutions using vanilla-PINNs with Dirichlet BCs}

We illustrate the use of our PINN-solver on two cases, i.e. for the two combinations of $(a_1, a_2, a_3)$ parameters that are
$(1, 0, 1)$ and $(1, 0 , -1)$. The other physical parameters are $L = 3$ and $c = 0.8$.

We have chosen $30$ training data points per boundary layer (i.e. $N_{data} = 120$) with a random distribution, as one can
see in Fig. 17 (with red dots on the boundaries) where the predicted solution are plotted for the two cases.
The exact solution is used to prescribe the training data values as done for the previous equations.
For the collocation data set, $N_{data} = 700$ points are generated inside the integration domain
using a pseudo-random distribution (i.e. latin-hypercube strategy) as one can see with blue dots. 
The evolution of the loss functions with the training epochs
that is reported in Fig. 18  for the first case,  illustrates the convergence toward the predicted solution. Note that the training is stopped after $60000$ epochs.
We have chosen a network architecture having $7$ hidden layers with $20$ neurons
per layer (i.e. corresponding number of $2601$ learning parameters for the neural network), and a fixed learning rate of $ l_r = 3 \times 10^{-4} $.
For such Helmholtz PDE, the PDE loss function is taken to be $  \mathcal{F} =   u_{ xx} +  u_{ zz} + c^2 u$.

The error distribution at the end of the training for the first case
is plotted in Fig. 19  exhibiting a maximum absolute error of order $0.004$, which also
corresponds to a similar maximum relative error of order $0.002$ (the maximum magnitude solution value being of order two).
Note that, again the predicted PINNs solution and associated error distribution are obtained using a third set of points (different from the collocation points)
that is taken to be a uniform grid of $100 \times 100$ points here. In this way, once trained, the network allows to predict the solution quasi-instantaneously
at any point inside the integration domain, without the need for interpolation (as done for with classical tabular solutions).
One must also note that the error is higher near the boundary due to the higher gradient of the
solution and to the coexistence of data/collocation points in these regions.

The precision of PINN-solver is very good but less than more traditional methods (like in finite-element codes for example).
This is a general property of minimization techniques based on gradient descent algorithms (Press et al. 2007, Baty 2023a).
However, a finer tuning of the network parameters together with the introduction of optimal combinations for weights of the partial
losses can generally ameliorate the results.

\subsection{More general Helmholtz problems}

Let us consider the Helmholtz equation (i.e. Eq. 17) with $c$ coefficient obeying $c^2 = (c_x^2 + c_y^2) \pi^2$, with $c_x$ and $c_y$
being integers. There may exist different $c_x, c_y$ values for which the equation is satisfied in an approximate way. Equivalently,
we may have, $(c_x^2 + c_y^2) \pi^2 = c^2 (1 \pm \epsilon) \simeq c^2$, with $\epsilon$ a small value parameter. In this case, a vanilla-PINNs solver
is expected to not converge well or to fail to predict the exact solution.

\section{Grad-Shafranov equations}

Another equation represents a second important issue for approximating magnetic equilibria of plasma in the solar corona.
The latter that is known as Gad-Shafranov (GS) equation, is used to model curved loop-like structures. It is also
used to approximate the magnetohydrodynamic (MHD) equilibria of plasma confined in toroidal magnetic devices that aim
at achieving thermonuclear fusion experiments like tokamaks.

The GS equation can be written as

      \begin{equation}
       -   \left [  \frac { \partial^2 \psi  } { \partial R^2 } +  \frac { \partial^2 \psi  } { \partial z^2 }  -  \frac { 1} { R}   \frac { \partial \psi  }  { \partial R }   \right ] = G (R , z,  \psi) ,
      \end{equation}
with a formulation using $( R , \phi, z)$ cylindrical like variables. 
The scalar function $\psi (R, z)$ is the desired solution allowing to deduce the poloidal magnetic field ${\bf B_p }$ (component in the $(R, z)$ plane) via
\begin{equation}
        {\bf B_p } =  \frac {1} { R}  \nabla \psi \times {\bf e_\phi } + \frac {F (\psi)} { R} {\bf e_\phi }
      \end{equation}
in the axisymmetric approximation, where $F (\psi)= R B_\phi$. The toroidal magnetic field component is $ {\bf B_\phi }   =  B_\phi {\bf e_\phi }$
oriented along the toroidal unit vector ${\bf e_\phi }$ (perpendicular to the poloidal plane).
The right hand side source term $G (R , z,  \psi)$ includes a thermal pressure term
and a second term involving a current density, and must be specified in order to solve the equation 20 (see below).

The elliptic differential operator (left hand side of equation ) can be rewritten, as multiplying the equation by $R$, leads to the following residual form
   \begin{equation}
       \mathcal{F} =  \left [  R \frac { \partial^2 \psi  } { \partial R^2 } + R  \frac { \partial^2 \psi  } { \partial z^2 }  -    \frac { \partial \psi  }  { \partial R }   \right ] + R G (R , z,  \psi) = 0 ,
      \end{equation}
that is effectively used in our PINN solver.

\subsection{Example of Soloviev equlibrium: the drop-like structure}

\begin{figure}[h]
    \centering
    \includegraphics[width=0.49\textwidth]{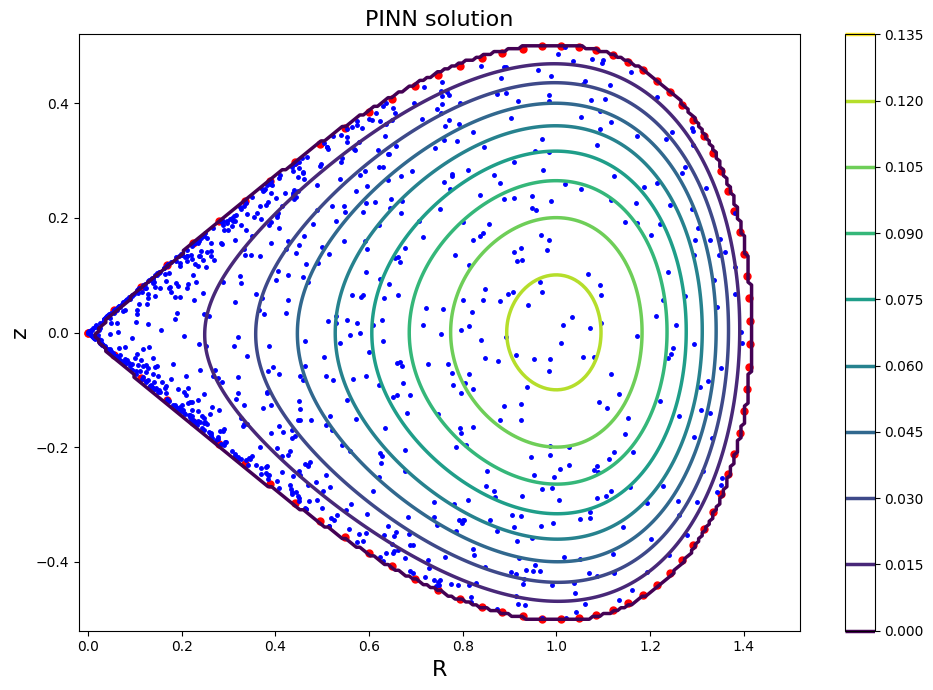}
    \includegraphics[width=0.49\textwidth]{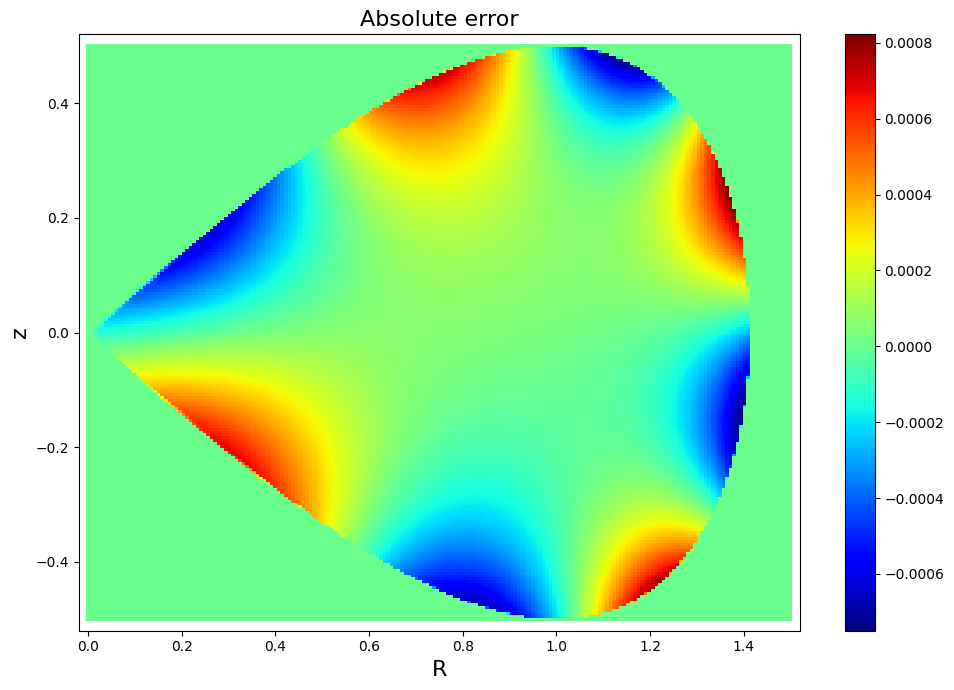}
    \caption{ Predicted solution (left panel) and absolute error distribution (right panel) using colored iso-contours, for the drop-like solution of
    GS equation and Soloviev equilibrium. The spatial locations for training and collocation data sets used on the boundary and interior domain respectively are
    indicated with red and blue dots respectively.
      }
\end{figure}

Exact analytical solutions called Soloviev solutions are of particular importance for approximating the general solutions relevant
of tokamaks and other variants of such toroidal magnetic devices (Soloviev 1975). The latter are obtained by taking relatively simple expressions
for the source term $G (R, z)$.

For example, assuming $G = f_0 (R^2 + R_0^2)$ leads to the exact analytical solution (see Deriaz et al. 2011)
   \begin{equation}
       \psi =    \frac { f_0 R_0^2 } { 2 }      \left [  a^2 - z^2 - \frac { R^2 - R_0^2 } { 4 R_0^2}   \right ] ,
      \end{equation}
with a simple boundary condition $\psi = 0$ on a closed contour $\partial \Omega$ defined by
   \begin{equation}
       \partial \Omega =    \left [   R = R_0  \sqrt  { 1 + \frac { 2 a \cos (\alpha) } {R_0}   }   , z = aR_0, \alpha =  [0 : 2 \pi] \right ] ,
      \end{equation}
where $R_0$, $a$, and $f_0$ are parameters to be chosen.
As can be seen below, the integration domain $\Omega$ bounded by $\partial \Omega$ has a funny drop-like form with an $X$-point 
topology at $z = R = 0$, as $\frac { \partial  \psi} { \partial  z} = \frac { \partial  \psi} { \partial  R} = 0$ at this point.

Here, we present the results obtained with our PINN solver using parameter values, $f_0 = 1, a = 0.5$, and $R_0 = 1$.
 The network architecture is similar to the arcade problem where $7$ hidden layers with $20$ neurons per layer were chosen, which consequently represent a number of $2601$ trainable parameters. We have used $80$ training data points (i.e. $N_{data}  = 80$) with a distribution based on a uniform $\alpha$ angle generator, and $N_c = 870$ collocation points inside the integration domain. Contrary to the case reported in Baty \& Vigon (2024), the distribution of collocation points is obtained using a pseudo-random generator with an additional concentration close to the $X$-point
 in order to get a better predicted solution there. The results are obtained after a training process with a maximum of $60000$ epochs.

\subsection{Examples of Soloviev equilibria: toroidal fusion devices}

\begin{figure}[h]
    \centering
    \includegraphics[width=0.43\textwidth]{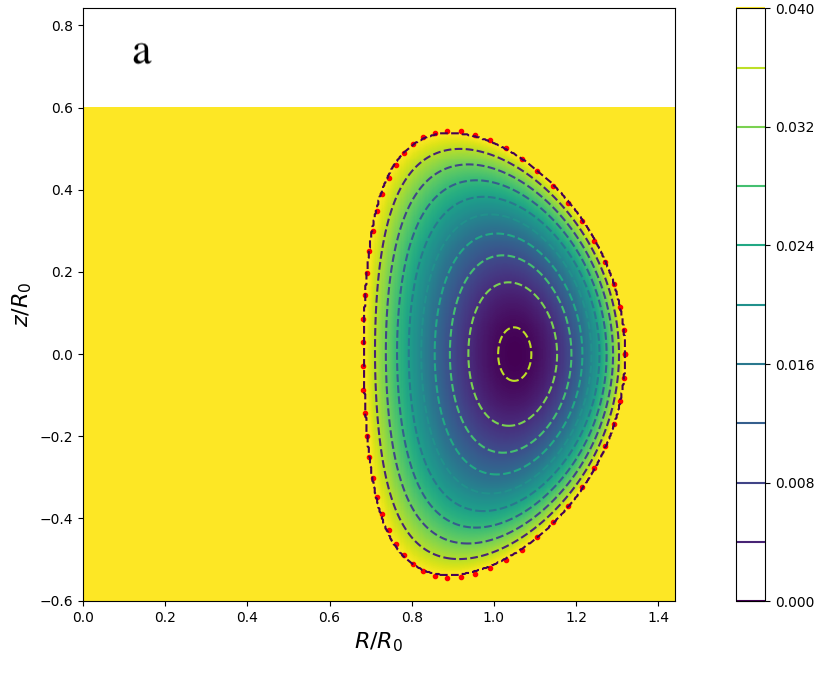}
    \includegraphics[width=0.43\textwidth]{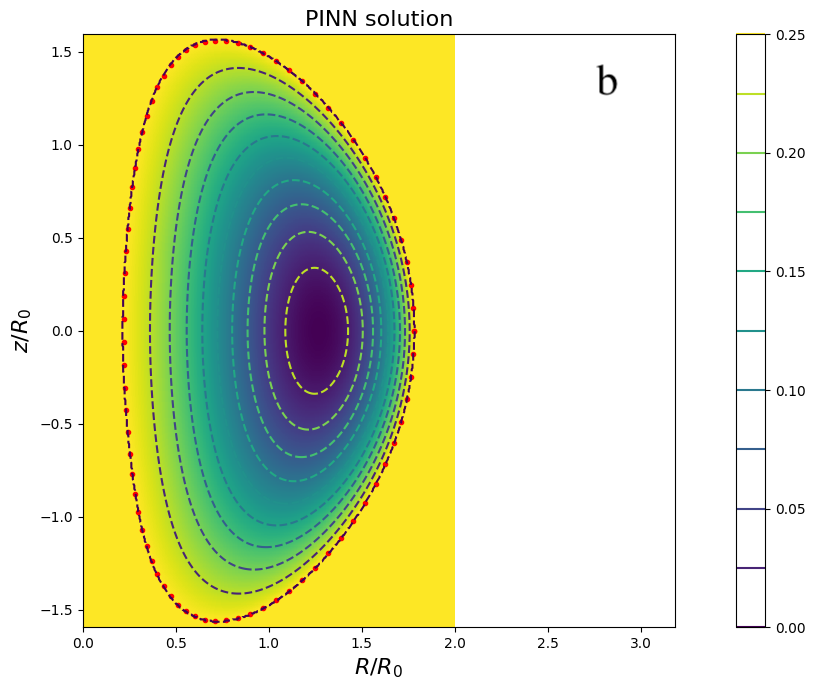}
    \includegraphics[width=0.43\textwidth]{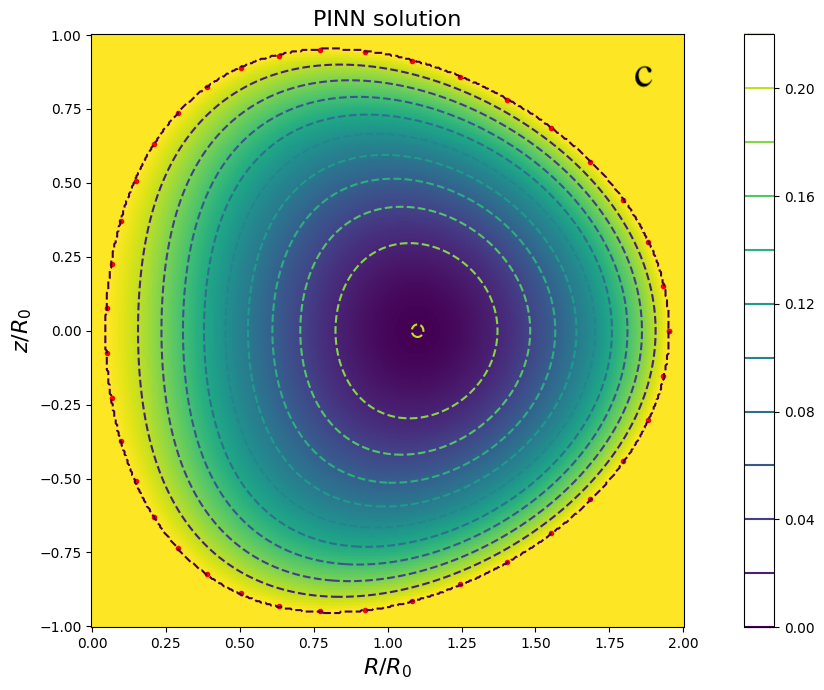}
    \caption{ Predicted solutions (colored iso-contours of $\psi$) for tokamak ITER-like (a panel), spherical tokamak NSTX-like (b panel), and spheromak-like (c panel)
    devices.  Only spatial locations of training data points where $\psi = 0$ (at the boundary) is imposed as soft (Vanilla-PINN) constraints are
    indicated with red dots.
      }
\end{figure}

In a similar way, a PINN solver can be used to predict Soloviev-like equilibria corresponding to the parametrization $G = A + (1 - A) R^2$
for the source term in the GS equation ($A$ being a dimensionless parameter).
The boundary of the integration domain where $\psi$ is expected to vanish is defined in a parametric way as
   \begin{equation}
       R =    1 + \epsilon \cos (\tau + \arcsin (\delta) \sin (\tau))        ,
             \end{equation}
and
 \begin{equation}
       z =    \epsilon \kappa \sin(\tau)  ,
             \end{equation}
where geometric parameters $\epsilon = a/R_0$ (inverse aspect ratio), $\kappa$ (the elongation), $\delta$ (the triangularity) are
introduced. The remaining parameter $\tau$ is an angle varying continuously in the range $[0, 2 \pi]$. Note that, as described by Cerfon et al. (2011)
the exact solutions are only approximately analytic. However, this parametrization presents the advantage to model Soloviev equilibria
for a collection of different toroidal devices (i.e. not only tokamaks for example) as illustrated below. The geometric parameters
also closely correspond to measured values from the experiments.

Using PINNs solvers similar to the previously described ones, predicted solutions corresponding to different toroidal devices
easily computed. Indeed, choosing the parameters combination, $\epsilon = 0.33$, $\kappa = 1.7$, and $\delta = 0.32$, together
with $A = -0.155$, allows to model magnetic equilibria representative of the ITER tokamak configuration. A second combination
with $\epsilon = 0.78$, $\kappa = 2$, and $\delta = 0.35$, together
with $A = 1$, allows to model magnetic equilibria representative of the NSTX spherical tokamak. And, a third combination
with $\epsilon = 0.95$, $\kappa = 1$, and $\delta = 0.2$, together
with $A = 1$, allows to model magnetic equilibria representative of a spheromak. 
The results obtained are plotted in Fig. 21. The particular choice for these parameter values is explained in details in Cerfon et al. (2011)
The choice for the neural network architecture and the numerical parameters for the gradient descent algorithm is similar
to what has been done previously.
Finally, note that a similar PINN solver tested on the same toroidal fusion devices have been developed by Jang et al. (2024).

\subsection{Other examples of more general equilibria in a rectangular domain}

\begin{figure}[h]
    \centering
    \includegraphics[width=0.32\textwidth]{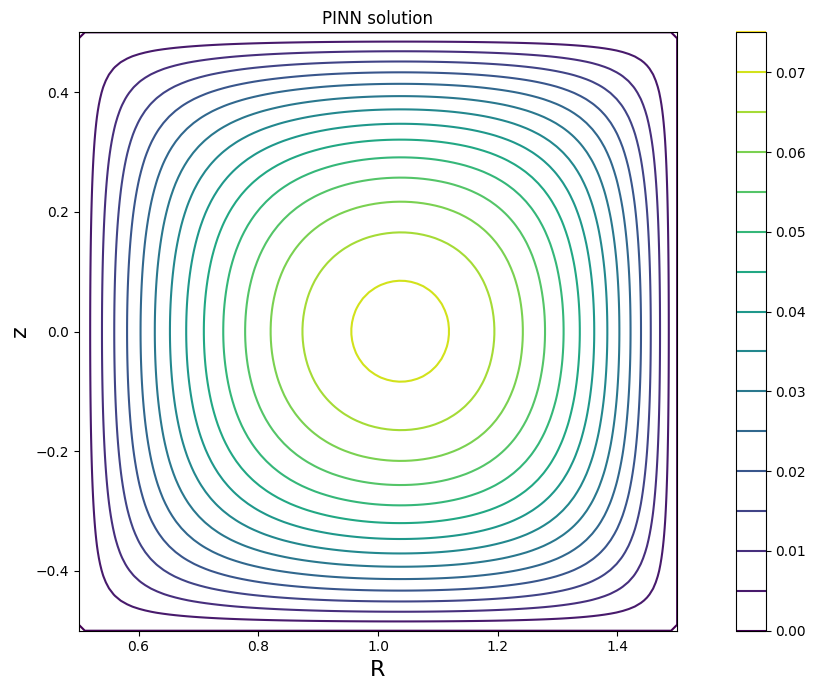}
    \includegraphics[width=0.33\textwidth]{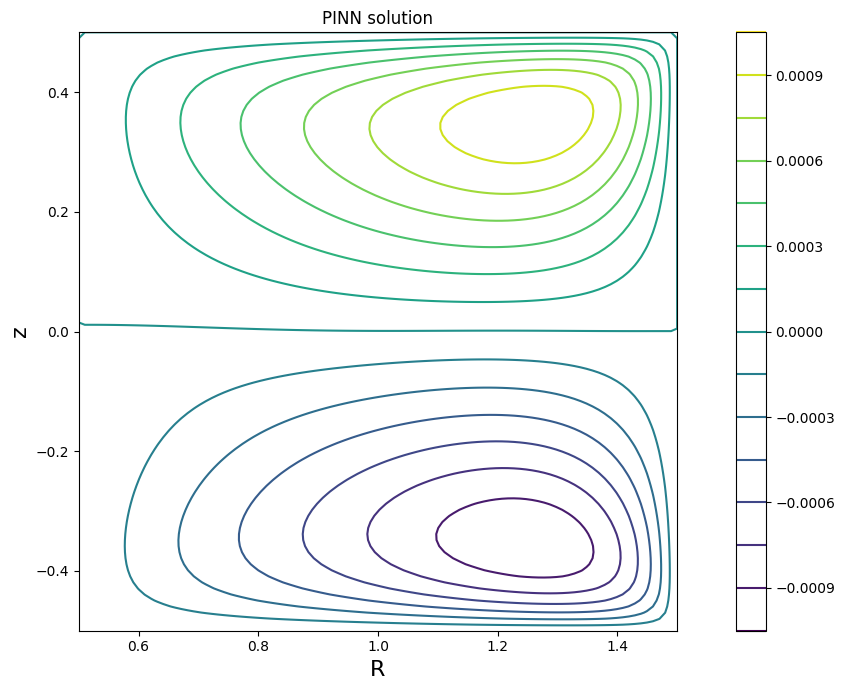}
    \includegraphics[width=0.33\textwidth]{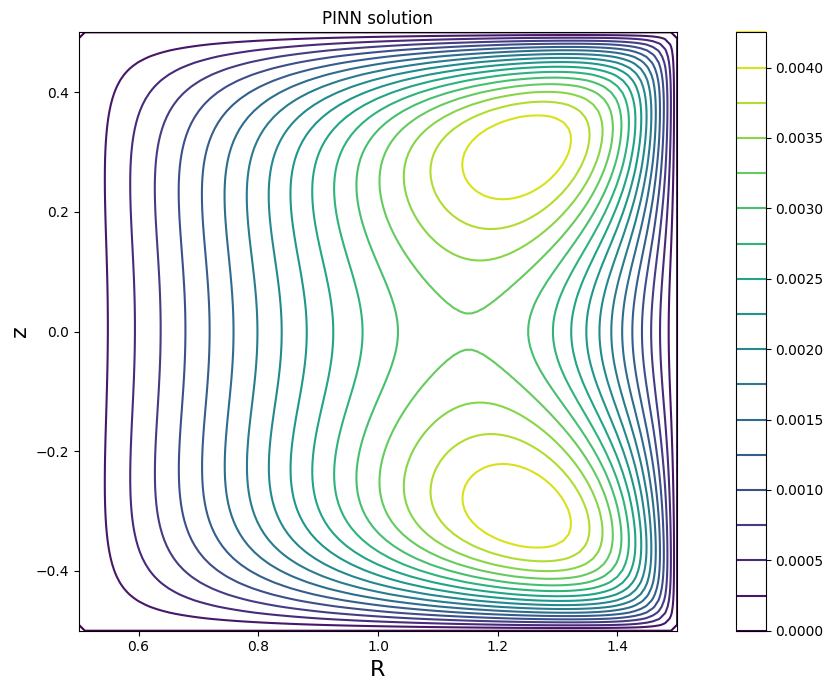}
    \caption{ Predicted solutions (iso-contours of $\psi$) of GS equation in a rectangular domain with three source terms,
    $G = 1$, $G = R^2 z^3$, and $G = R^3 z^2$ in left, middle, and right panel respectively. Hard constraints are used
    to impose Dirichlet BCs.
      }
\end{figure}

\begin{figure}[h]
    \centering
    \includegraphics[width=0.45\textwidth]{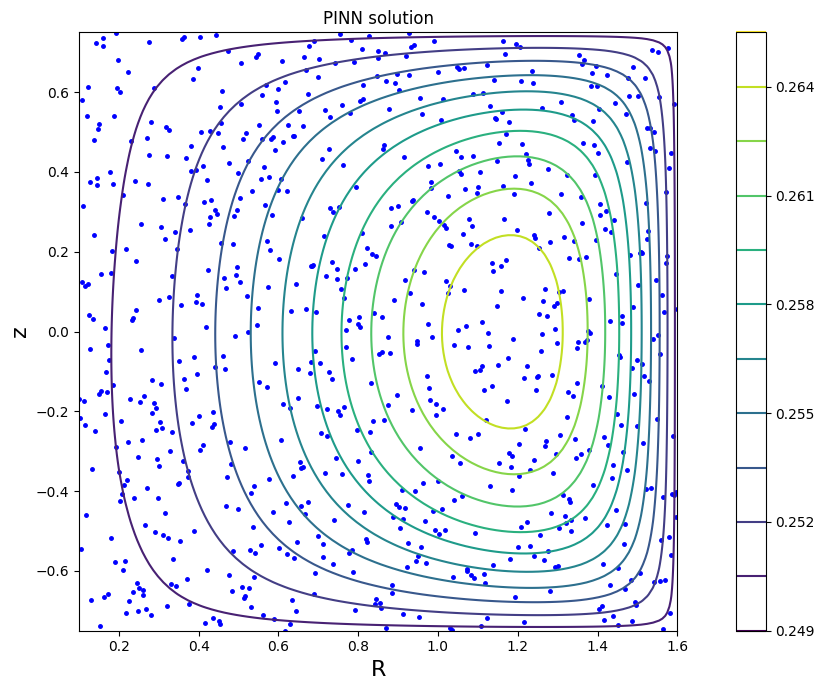}
    \caption{ Predicted solutions (iso-contours of $\psi$) of GS equation in a rectangular domain with a
    non linear source term $G$ (see text). The location of collocation points are indicated with blue dots. There is no training data points
    as hard constraints (Lagaris method) are used to impose Dirichlet BCs.
      }
\end{figure}

As a simplification of the boundary, solutions of GS equilibria can be computed in a rectangular domain (see Itagaki et al. 2004).
where the homogeneous solution $\psi = 0$ is imposed.
Source terms of different forms can be studied like $G = R^l z^m$ with $m$ and $n$ being positive integer parameters.

A rectangular domain $\Omega = [0.5, 1.5] \times [-0.5, 0.5]$ is taken.
Note that, for this particular problems, the use of hard constraints (with Lagaris like BCs) leads to better results compared to 
the use of vanilla-PINNs. Indeed, following the specification proposed by Lagaris (1998) and as expressed in subsection 3.2,
it is preferable to write 
   \begin{equation}
       \psi_\theta =    (R - 0.5) (R -1.5) (z - 0.5) (z + 0.5)   \psi_\theta^*   ,
             \end{equation}
where $ \psi_\theta^*$ is the output of the neural network. In this way, the predicted solution $ \psi_\theta$ automatically
vanishes at the boundary and the training data set is not needed. 
Using PINNs solvers similar to the previously described ones, predicted solutions for three different source terms are plotted
in Fig. 22. 

\
In order so show that PINN solver can also be used for non linear source terms $G$, we consider a case taken from Peng et al. (2020) with
$G = 2 R^2 \psi [c_2 (1 - \exp( -\psi^2/\sigma^2) + 1/\sigma^2 (c_1 + c_2 \psi^2) \exp(-\psi^2/\sigma^2)]$ with
$\sigma^2 = 0.005, c_1 = 0.8$, and $c_2 = 0.2$ in a domain $\Omega = [0.1, 1.6] \times [-0.75, 0.75]$. The results
obtained with a PINNs solver using hard constraints are plotted in Fig. 3. In this case, the imposed Dirichlet condition value used
is now $ \psi = 0.25$. Consequently, we also use
   \begin{equation}
       \psi_\theta =    (R - 0.1) (R -1.6) (z - 0.75) (z + 0.75)   \psi_\theta^*   ,
             \end{equation}
As there is  no exact solution available, this is not possible
to evaluate the error. However, our solution compares well with the solution computed in Peng et al. (2020).

\section{Lane-Emden equations}

The Lane-Emden (LE) equations are widely employed in astrophysics and relativistic mechanics. Before focusing on a precise
case taken form astrophysics, it is instructive to introduce a more general mathematical form.

\subsection{A mathematical example}

\begin{figure}[h]
    \centering
    \includegraphics[width=1.\textwidth]{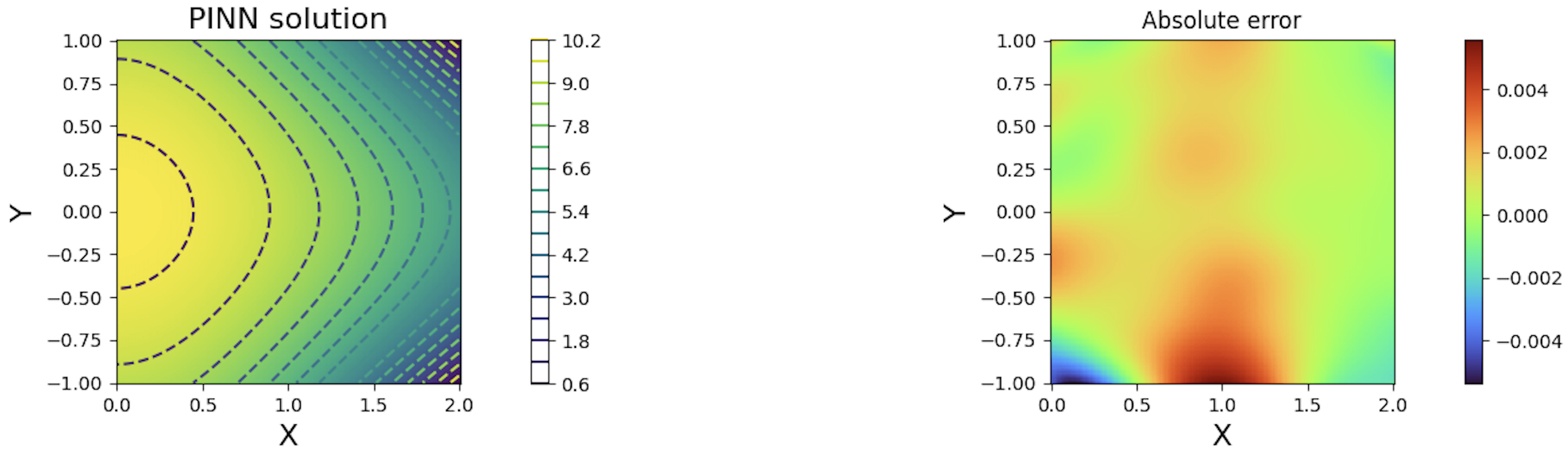}
     \includegraphics[width=1.\textwidth]{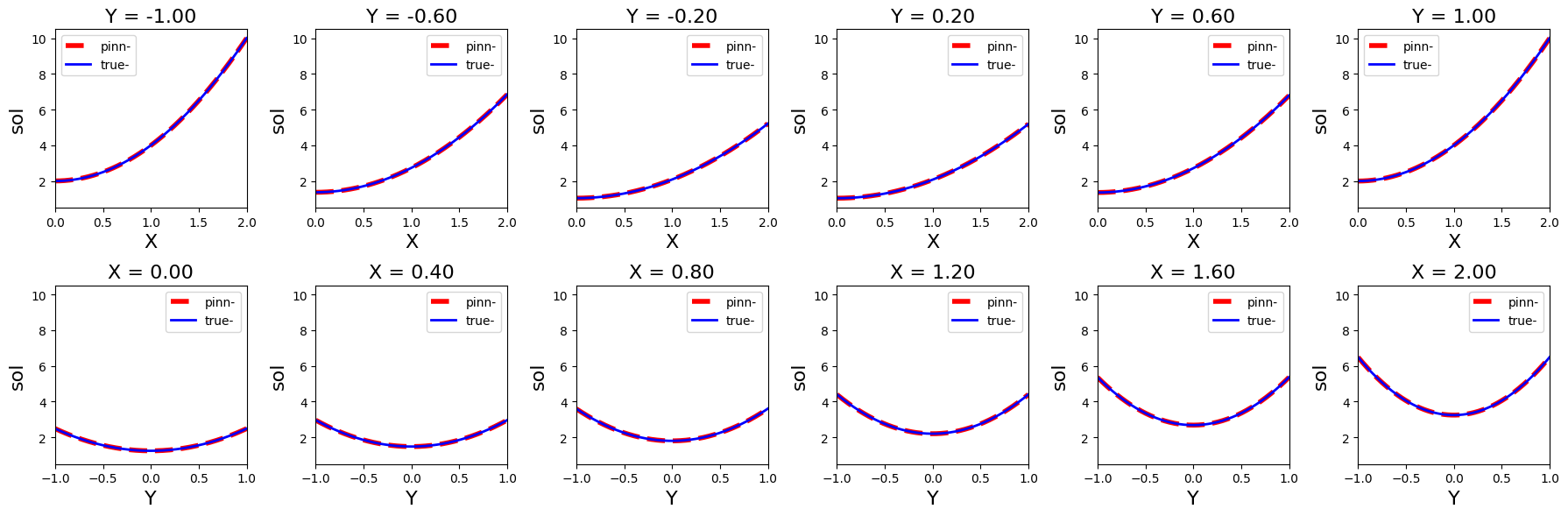}
    \caption{ Predicted solution (iso-contours of $\psi$ in top-left panel) of mathematical LE equation in a rectangular domain
    with Cauchy condition imposed at two boundaries ($x = 0$, and $x = 2$). The absolute error is plotted
    in top-right panel. One dimensional cuts (at given $y$ and $x$ values) show the predicted solution versus the
    exact one.
      }
\end{figure}

\begin{figure}[h]
    \centering
    \includegraphics[width=1.\textwidth]{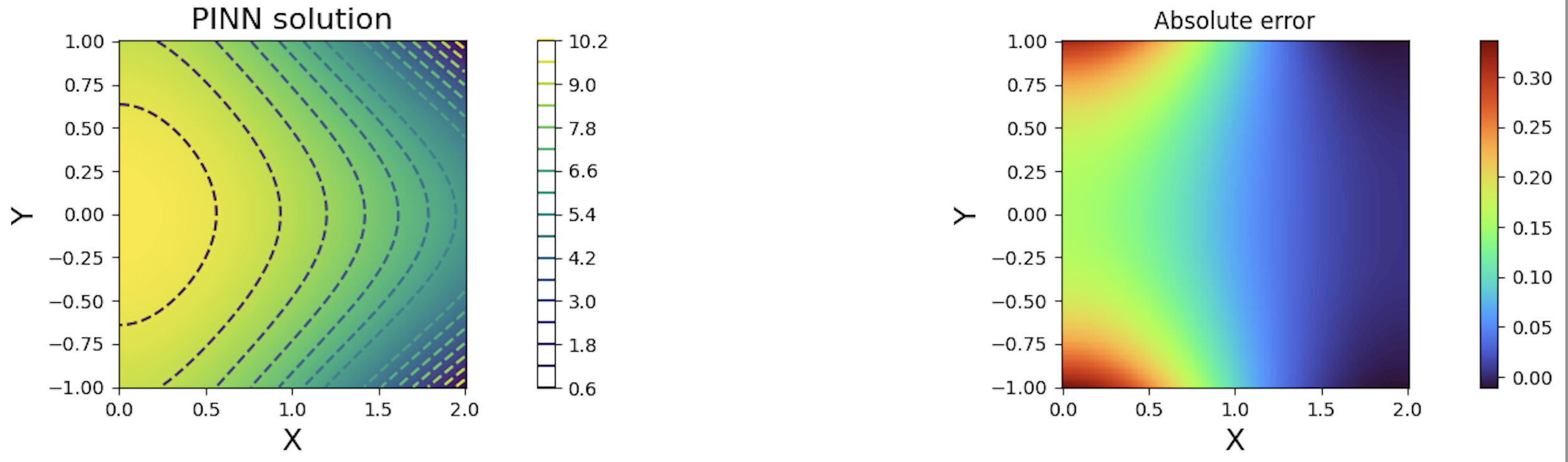}
     \includegraphics[width=1.\textwidth]{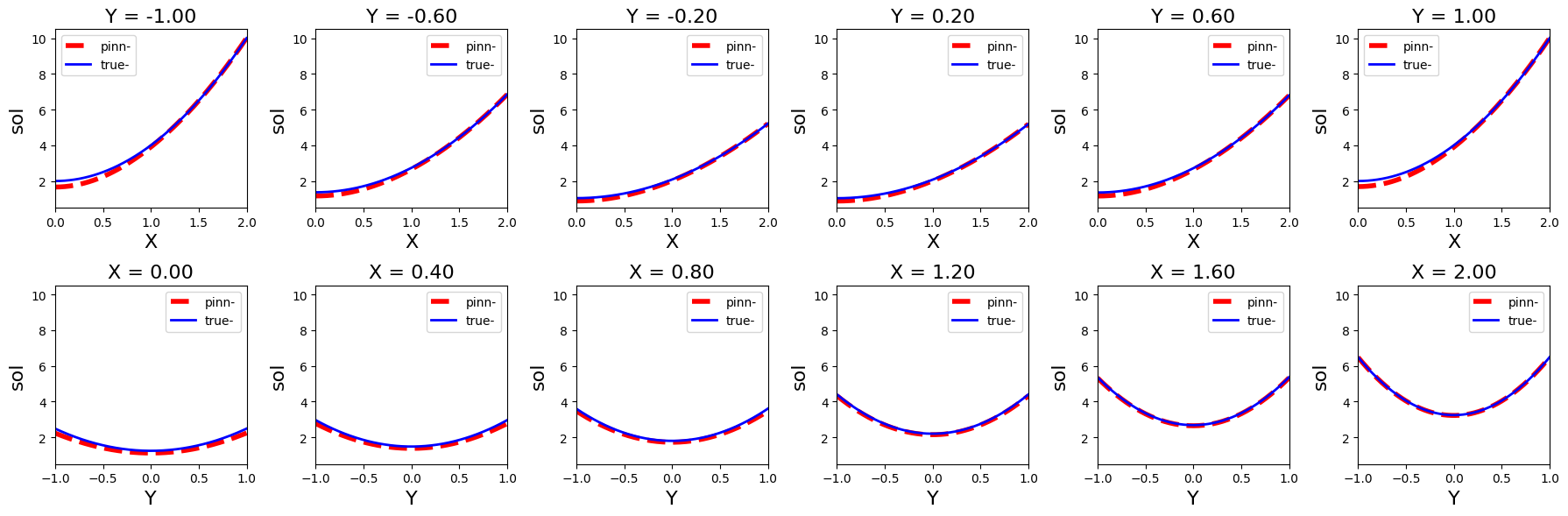}
    \caption{ Same as in previous figure, but using Cauchy condition at only one boundary (i.e. at
     $x = 2$).
      }
\end{figure}

\begin{figure}[h]
    \centering
    \includegraphics[width=1.\textwidth]{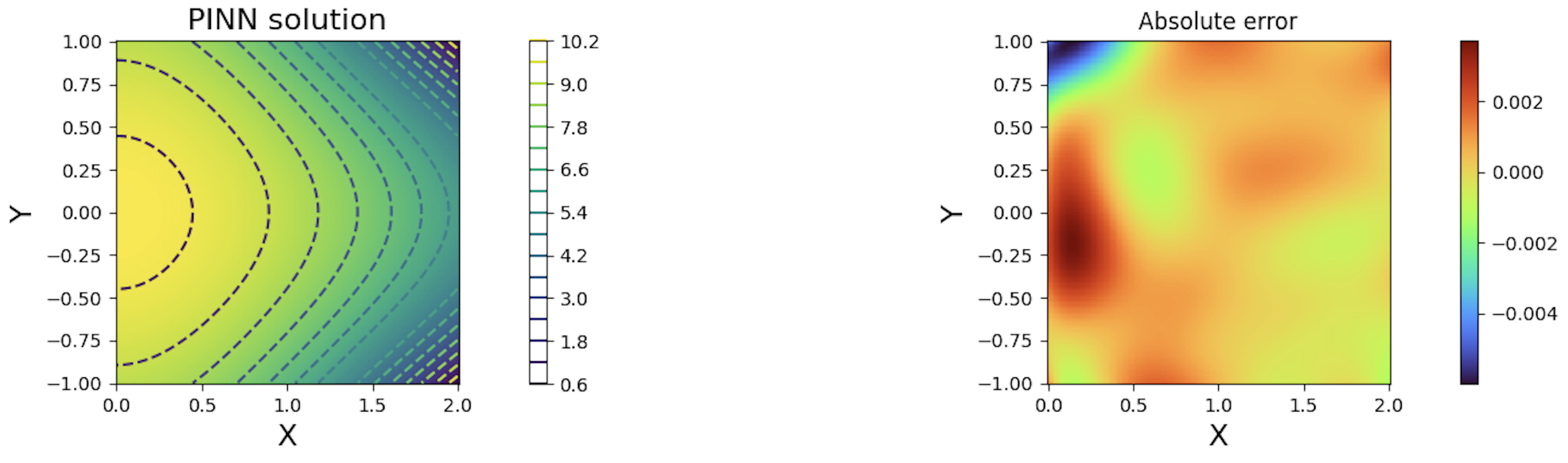}
     \includegraphics[width=1.\textwidth]{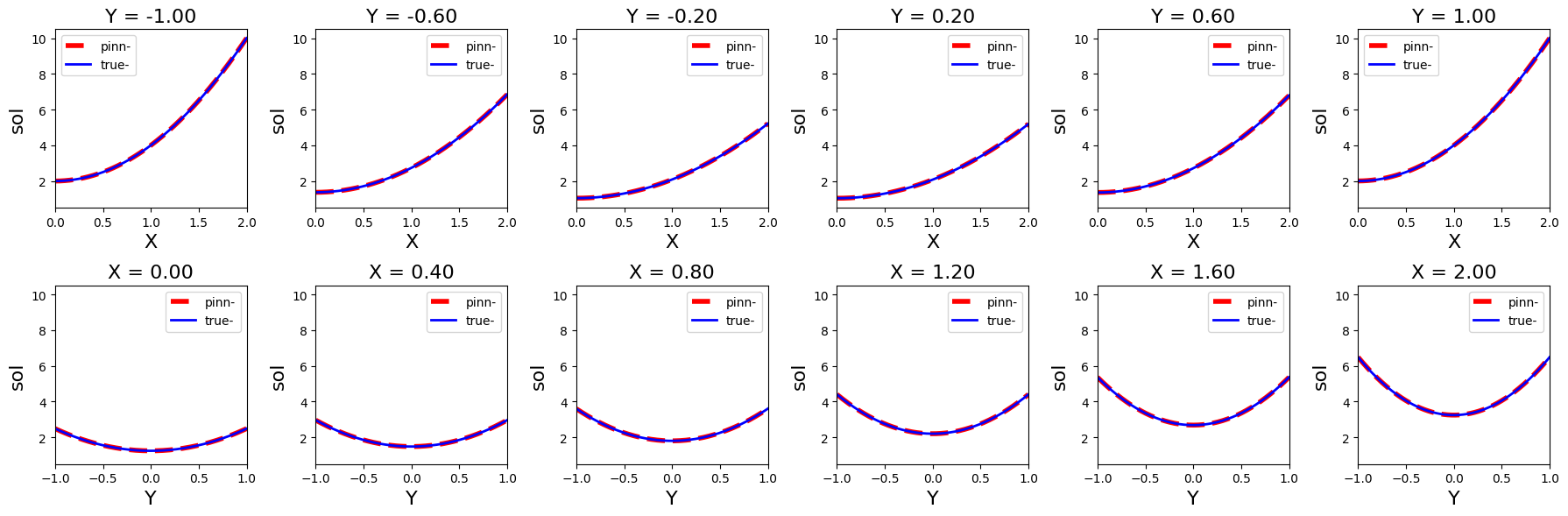}
    \caption{ Same as in previous figure, but using Cauchy condition at only one boundary (i.e. at
     $x = 2$) and Neumann condition at the two boundaries $y  \pm 1$.
      }
\end{figure}

Following previous papers (Bencheikh 2023 and references therein), we can consider the following two-dimensional differential
form:
   \begin{equation}
      u_{xx} + \frac { \alpha } { x}  u_x + u_{yy} + \frac { \beta } { y}  u_y  +  f(u, x, y) = 0 ,
          \end{equation}
where second order derivatives ($u_{xx}$ and $u_{yy}$), and first order derivatives ($u_{x}$ and $u_{y}$) of the desired
solution are involved. The two scalars $\alpha$ and $\beta$ are real shape parameters, $f(x, y)$ is a given scalar 
source function, and integration is done on a cartesian $(x, y)$ domain.

The particularity of Lane-Emden equation lies in the singularities at $x = 0$ and $y = 0$, which must be 
overcome by numerical integration method. As can be seen below, solvers based on PINNs algorithm are
excellent choice as classical discretization is not needed.
We also focus on the search for solutions in a rectangular domain with Cauchy-like conditions imposed
at one or two of the four boundaries. The conditions at the other boundaries are assumed free or of
Neumann type. Indeed, such problems are representative of physical examples in astrophysical context (see second
subsection in this section).
            
As an example, we take a case with $ \alpha =  \beta = 2$, and $f = - 6 ( 2 + x^2 + y^2)$. The integration domain is
$\Omega = [0, 2] \times [-1, 1]$. The exact solution can be checked to be $ u = (1+x^2) (1 + y^2)$.
The residual form used to evaluate the PDE loss function is taken to be
   \begin{equation}
    \mathcal{F} =  x y u_{xx} + \alpha y  u_x + x y u_{yy} + \beta x u_y  +  x y f(u, x, y) = 0 ,
          \end{equation}

Using a PINNs solver similar to previously described ones, we can predict the solution for three distinct
problems involving Cauchy-type condition at (at least) one boundary.
Indeed, when we imposed Cauchy condition (i.e. using the exact solution and also the its perpendicular derivative)
at the two boundaries $x = 0$ and $x = 1$, the predicted PINN solution (plotted in Fig. 24) shows a rather good
agreement with the exact solution inside the domain, as the maximum absolute error is $0.004$.
However, when the Cauchy condition is imposed only at one boundary, i.e. at $x = 2$, the accuracy is
significantly deteriorated compared to the previous case (see Fig. 25 exhibiting a maximum absolute error of
$0.30$). This error is obviously mainly localized at the opposite boundary.
In astrophysical context, the prescription of the perpendicular derivative at the two other boundaries (i.e. at
$y  \pm 1$ can be a a reasonable hypothesis. Consequently, when we impose Cauchy condition at $x = 2$
in addition to Neumann conditions at $y  \pm 1$ , one can see that the predicted PINN solution is improved
again. The latter result is clearly in Fig. 26, where the maximum absolute error is comparable to the first case
with Cauchy condition imposed at the two opposite boundaries.

\subsection{A physical example}

\begin{figure}[h]
    \centering
    \includegraphics[width=0.4\textwidth]{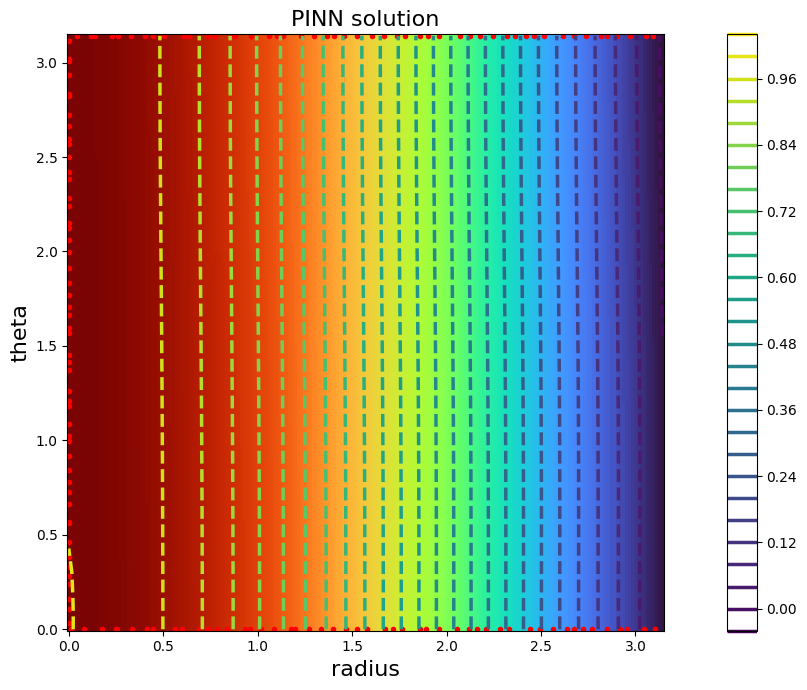}
     \includegraphics[width=0.4\textwidth]{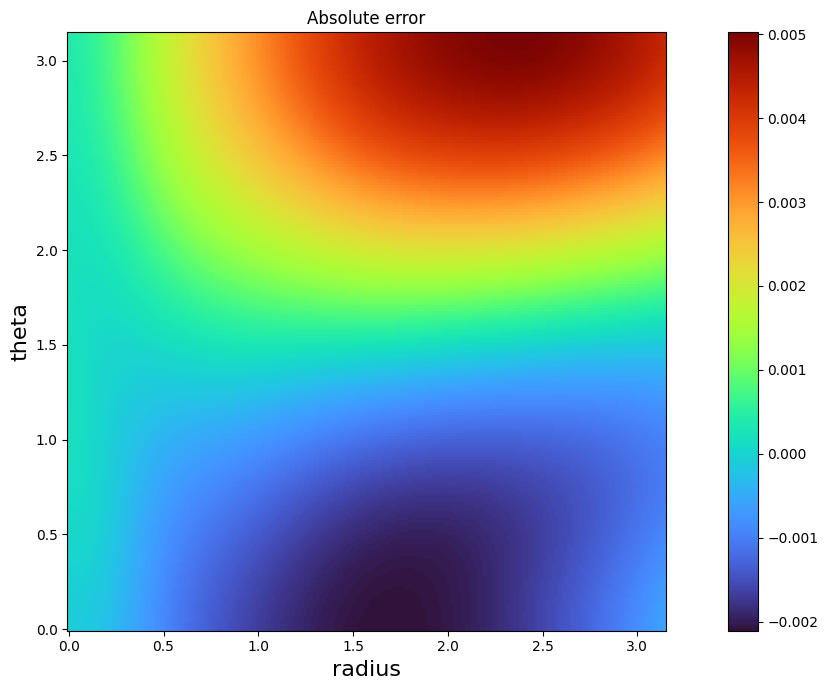}
    \caption{ Predicted PINNs solution (left panel) and associated absolute error distribution (right panel) obtained from the physical Lane-Emden problem for $n = 1$ and $\omega = 0$ in the ($r, \theta$) plane.
      }
\end{figure}

\begin{figure}[h]
    \centering
    \includegraphics[width=0.44\textwidth]{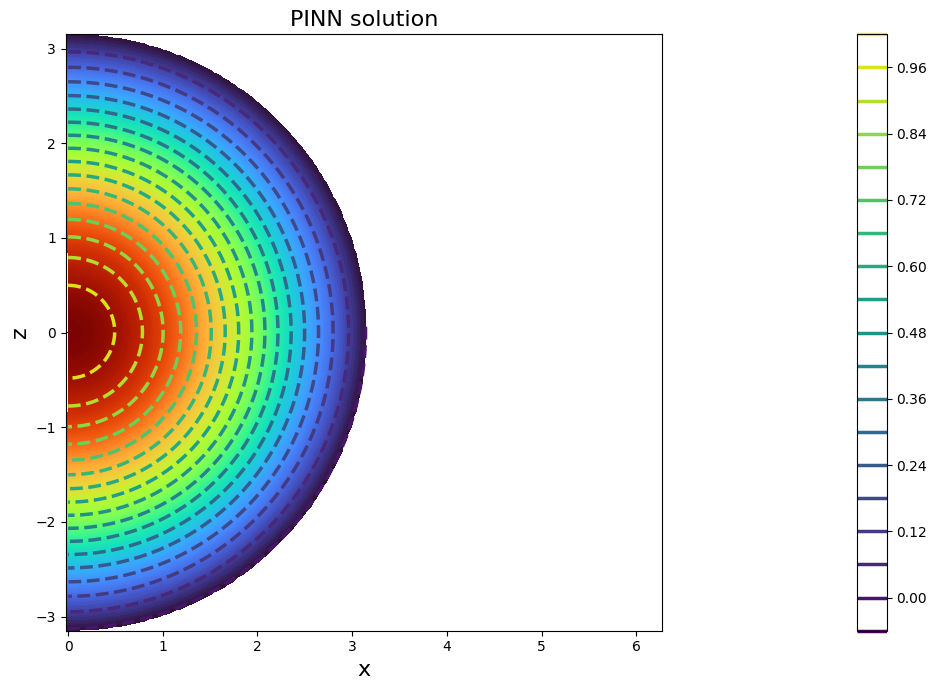}
     \includegraphics[width=0.34\textwidth]{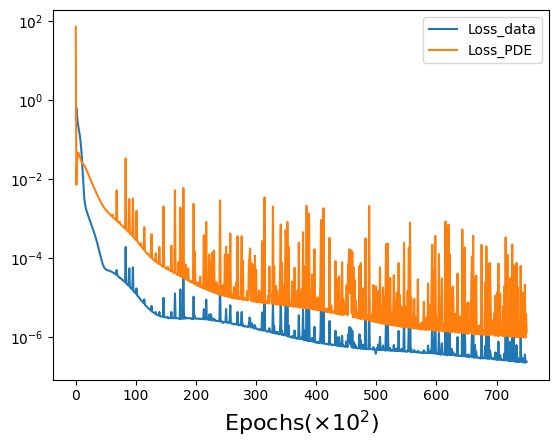}
    \caption{ (Left panel) Predicted PINNs solution obtained from the physical Lane-Emden problem for $n = 1$ and $\omega = 0$ (see previous figure) plotted in the associated ($z, x$) plane.
    (Right panel) Evolution of the losses during the training process.
      }
\end{figure}

\begin{figure}[h]
    \centering
    \includegraphics[width=0.9\textwidth]{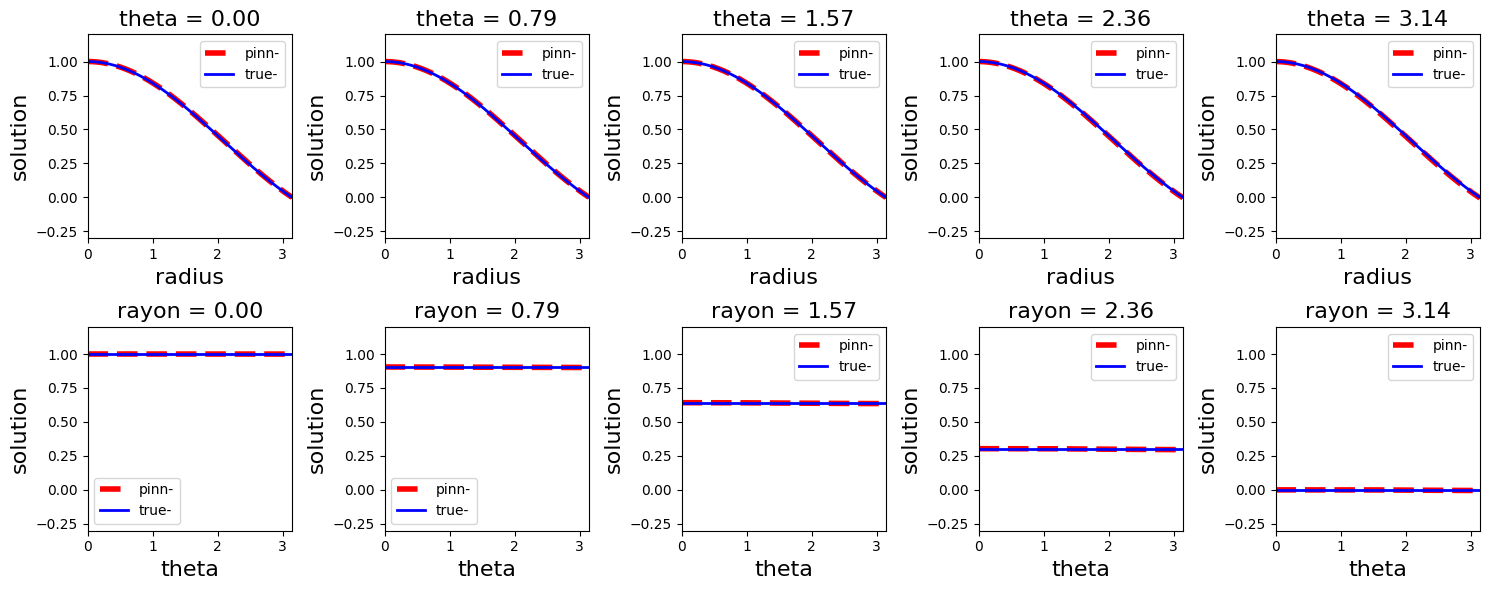}
        \caption{ One dimensional cuts (at given $r$ and $\theta$ values) showing the predicted solution versus the
    exact one.
      }
\end{figure}

In this subsection, we consider the problem related to the internal structure of a rotating self-gravitating gaz sphere in hydrostatic 
equilibrium. In astrophysics, this is an approximation for different stars when a polytropic relation between the thermal
pressure $P$ and the gaz density $\rho$ is taken at any point inside the star, i.e. $P = K \rho^{(1 + 1/n)}$ where $n$ is a polytropic index having
different values according to the type of star studied and $K$ is a constant.

Using two physical equations, a first one related to the mass conservation (continuity equation) at any radius, and a second one
related to equilibria between different forces (thermal pressure gradient force, gravitational force, and centrifugal force), one
can recast the problem into a single second order differential equation that is the precisely the Lane-Emden equation.
The latter can be written as
      \begin{equation}
      \Delta \psi + \psi^n = \omega ,
          \end{equation}
where 
$\Delta$ is the two dimensional spherical Laplacian operator $\Delta = \frac {1 } { r^2} \frac { \partial} {  \partial r} (r^2   \frac {\partial } {  \partial r} ) + 
 \frac {1 } { r^2 \sin \theta}  \frac { \partial} {  \partial \theta} (\sin \theta  \frac   {\partial } {  \partial \theta}  )$ and $\omega$ is a constant representative of
 the rotation of the star (assumed to be uniform). Physically speaking,  $\psi$ is related to the mass density (see Baty 2023). For example, in the
 $n = 1$ case, it is exactly the mass density normalized to the value at the center.
 This is a dimensionless equation depending on two spatial coordinates (spherical geometry) with $r = R/R_c$ ($R_c$ being
the star radius) and with $\theta$ the co-latitude angle varying between $0$ and $\pi$. The problem is assumed axisymmetric and then does not depend on the remaining
spherical angle. The latter can be also developed in the equivalent form (that is closer to the previous mathematical equation),
   \begin{equation}
      \frac {\partial^2 \psi} {\partial r^2 }  + \frac {2 } { r}   \frac {\partial \psi} {\partial r} +  \frac {1 } { r^2}   \frac {\partial^2 \psi} {\partial \theta^2 }  +      
      \frac {1 } { r^2 \tan \theta}  \frac {\partial \psi} {\partial \theta} +  \psi^n = \omega .
          \end{equation}
However, contrary to the mathematical form (see Eq. 28), this physical form displays an obvious asymmetry between the two coordinates
$r$ and $\theta$.

We can solve the previous LE equation in the particular case $n = 1$ without rotation (i.e. $\omega = 0$). In this case, an analytical solution exists
that is purely radial as $\psi = \sin(r)/r$. A PINN solver is develped using a PDE loss function based on the residual equation,
   \begin{equation}
  \mathcal{F} =  r^2  \frac {\partial^2 \psi} {\partial r^2 }  +  2r  \frac {\partial \psi} {\partial r} +   \frac {\partial^2 \psi} {\partial \theta^2 }  +      
     \frac {1 } { \tan \theta}    \frac {\partial \psi} {\partial \theta} + r^2 \psi^n = 0 .
          \end{equation}
The BCs used in this problem are Cauchy condition on the axis $r = 0$, that are $\psi = 1$ and $\frac {\partial \psi} {\partial r} = 0 $. Neumann
BCs are also added at the two boundaries $\theta = 0, 2 \pi$, that are $\frac {\partial \psi} {\partial \theta} = 0 $. 
Using $1000$ collocation points randomly distributed in the domain $\Omega = [0, \pi] \times [0, 2 \pi]$ and $150$ training data points
at the three boundaries ($50$ per boundary), our PINNs solver is able to predict the exact solution 
with a precision similar to ones obtained for previous problems, as one can see in Figs 27-29. A total number of $70 000$ is used
for this example.

Solutions for other index values (i.e. $n$) can be also easily obtained in the same way. Note that, without rotation (i.e. $\omega = 0$), the solutions are purely
radial (see Baty 2023a). In principle this is not the case when the rotation factor is not zero, and a physically relevant solution should depend
on the angle $\theta$. Unfortunately, the latter solution cannot be obtained using the Lane-Emden equation solely as some extra conditions
are lacking (see Chandrasekhar \& Milne 1938).

\section{Parametric differential equation}

\subsection{Parametric solution for stationary advection diffusion in one dimension}

In order to illustrate a PINNs solver aiming at learning multiple solutions of a given problem, we consider a simple differential
equation residual (written in a residual form) as follows
\begin{equation}
 c \frac  {\partial u}   {\partial x} - \mu  \frac  {\partial^2 u}   {\partial x^2}  - 1 = 0 ,
 \end{equation}
where $u (x)$ is the desired solution in a one dimensional spatial domain with $x$ in the range $[0, 1]$. As $\mu$ is a dissipation coefficient (i.e. viscosity),
and $c$ is a constant coefficient analog to a velocity (taken to be unity for simplicity),
 the latter equation represents a steady-state advection-diffusion problem with the additional constant source term (i.e. unity).  An example of exact solution
 with boundary conditions $u (0) = u(1) = 0$ is
 \begin{equation}
  u (x) = x -  \left [\frac  { e^{(x-1)/\mu }  -   e^{-1/\mu }  }   { 1 -   e^{-1/\mu }  } \right ] .
 \end{equation}
 Such solution is known to involve the formation of singular layers (i.e. at $x = 1$) when the viscosity employed is too small.
As we are interested by learning the solutions at different viscosities with the same neural network, we can consider now variable $\mu$ values taken in the range $[0.1, 1.1]$.
A PINNs solver can thus be easily designed where the second neuron (see Fig. 5) corresponds now to $\mu$ values, and the desired solution must
be properly called now $u (x, \mu)$. The corresponding residual equation form is now
 \begin{equation}
   \mathcal{F} \left [ u(x, \mu), x, \mu  \right ]  = 0.
\end{equation}

   \begin{figure}[!t]
\centering
  \includegraphics[scale=0.40]{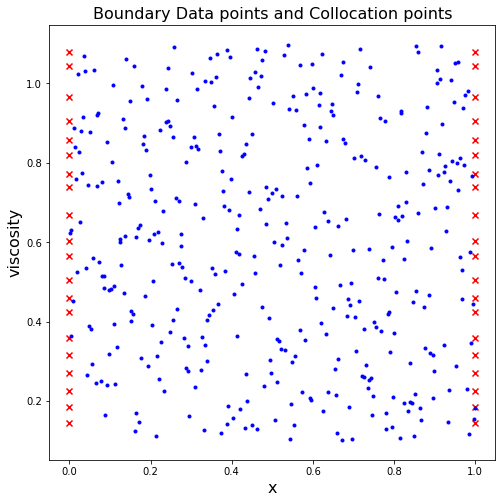}
  \caption{Scatter plot of the collocation points (blue dots) and of the training data (red crosses at the two boundaries $x=0, 1$ in the $(x, \mu)$ plane.
     }
\label{fig12}
\end{figure}   

   \begin{figure}[!t]
\centering
  \includegraphics[scale=0.38]{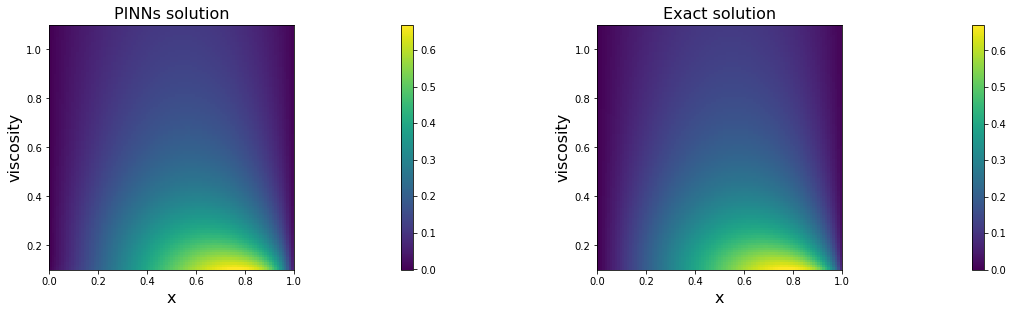}
  \caption{Colored iso-contours of the predicted PINNs solution $u (x, \mu)$ and exact solution in the left and right panel respectively.
       }
\label{fig13}
\end{figure}   

   \begin{figure}[!t]
\centering
  \includegraphics[scale=0.46]{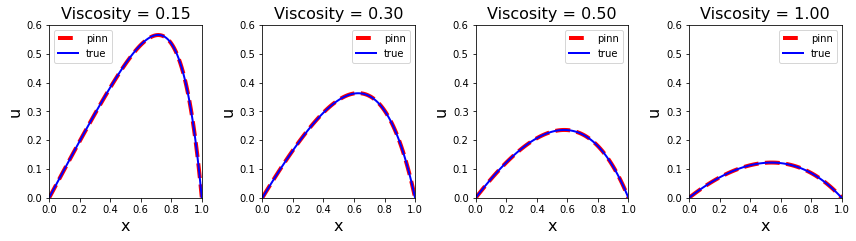}
  \caption{One-dimensional solution (red color) obtained for four different $\mu$ viscosity values situated in the range of learned values (see legend)
  compared with the exact analytical solution (blue color), corresponding to one-dimensional cut obtained from previous figure.
     }
\label{fig13}
\end{figure}   

We can generate random distributions of training boundary points (typically $20$ points at $x = 0$ and $20$ points at $x = 1$ corresponding to different
viscosity values in the range indicate above) and $400$ collocation points in the $(x, \mu)$ space $\Omega = [0, 1] \times [0.1, 1.1]$
as one can see in Fig. 30. The exact boundary values (i.e. with zero values in the present problem) are imposed at these boundaries in order to minimize the
training data loss function $L_{data}$, and the residual equation is evaluated on the collocation points for the variable viscosity in order to minimize
the physics-based loss function  $L_{PDE}$.
The results are plotted in Fig. 31 (iso-contours in the trained plane) and Fig. 32 (cuts at different viscosity value)
where one can see the predicted PINNs solution agrees very well with the exact one whatever the $\mu$ value, with an error similar to values reported for the previous problems.

\subsection{Stiff problems involving singular layers}

Advection-diffusion equations, as considered in this section, generally involve singular layers (e.g. at the domain boundary as shown in the previous problem at $x = 1$) that
are particularly stiff as the diffusion coefficient is small. More exactly, the relevant parameter is the Reynolds number $R_e = c/\mu$ which makes the problem
stiff when it is much greater than unity. This causes numerical difficulties with PINNs, as it is also the case when employing traditional numerical methods (Baty 2023c).

As an interesting potential use of PINNs method, one can use hard-PINNs solver where a formulation close to Lagaris one (Lagaris 1998) allows to well
capture the boundary layer solution at decreasingly small viscosity values.
More explicitly, this is obtained by using a trial function having a exponential decay as $\exp {(-x/\mu)} $ as it is the case
for the exact solution at the singular boundary. The latter algorithm is named semi-analytic PINN method (Gie et al. 2024).

\section{Inverse problem differential equation: parameter identification}

We consider the same advection-diffusion equation as taken in the previous section (see Eq. 33). However, we are now interested
in finding the unknown value for the viscosity coefficient $\mu$, from the knowledge of the equation together from a training data set
giving the solution at these points (i.e. at some $x$ values).

For example, using Eq. 35 (exact solution) we have generated some values (i.e. $40$) at different $x$ values (randomly distributed in the range $[0, 1]$)
for a given viscosity parameter that is $\mu = 0.15$. Moreover, in order to mimic data taken from measurements, we have added some random
noise with an amplitude of $0.01$ (see left panel of Fig. 33).

Our PINN solver is developed following the algorithm schematized in Fig.5. The aim is to use a PDE loss where the coefficient $\mu$ is not a priori known
and must be discovered/identified during the training process. The learning algorithm is similar to the non linear approximation scheme (Fig. 1) in
the sense that, now the training data set not only consists in the initial/boundary data. Moreover, a number of $33$ collocation points uniformly distributed
in the whole interval are chosen in order to evaluate the PDE loss.
As a result, at the end of the training (i.e. after $40000$
epochs), the solution predicted by the PINN solver is plotted in right panel of Fig. 33. The later is similar to the solution used to generate
the data using $\mu = 0.15$. We can also follow the estimation of the  $\mu$ parameter during the training as a function of the number of epochs.
The result is plotted in Fig. 34 (left panel), showing that it rapidly converge to a value close to the (exact) expected value of $0.15$
after $5000$ epochs. More precisely, a value that is slightly smaller is obtained due to the noise effect. This is confirmed by the result using
training data without noise, as one can see in right panel of Fig. 34.

   \begin{figure}[!t]
\centering
  \includegraphics[scale=0.44]{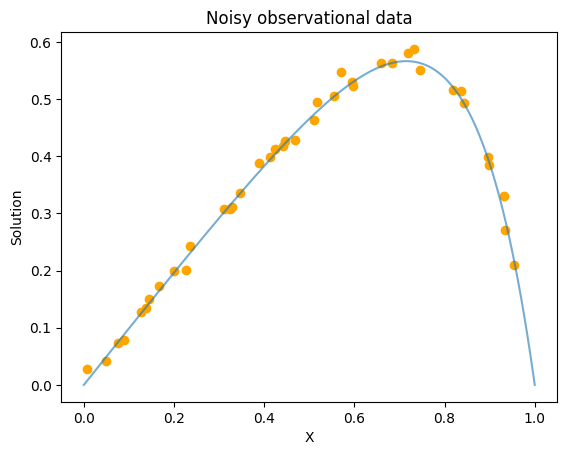}
   \includegraphics[scale=0.63]{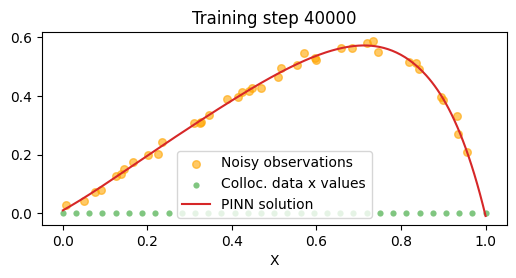}
  \caption{(Left panel) Set of data generated at different $x$ values for a viscosity parameter $\mu = 0.15$ from the expression Eq. 34 with additional
  random noise amplitude value $0.01$. (Right panel) Predicted PINNs solution obtained at the end of the training process (after $40000$ epochs).
     }
\label{fig12}
\end{figure}   

   \begin{figure}[!t]
\centering
    \includegraphics[scale=0.49]{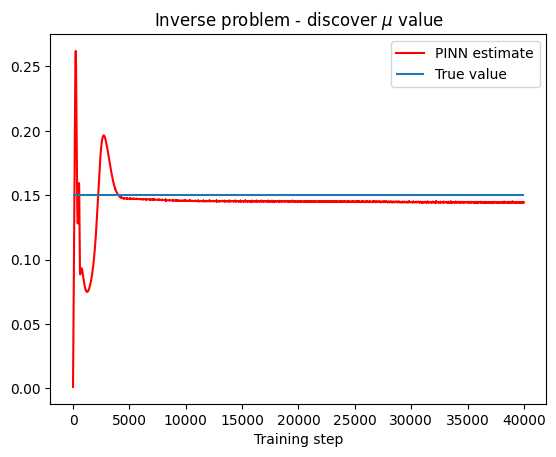}
   \includegraphics[scale=0.49]{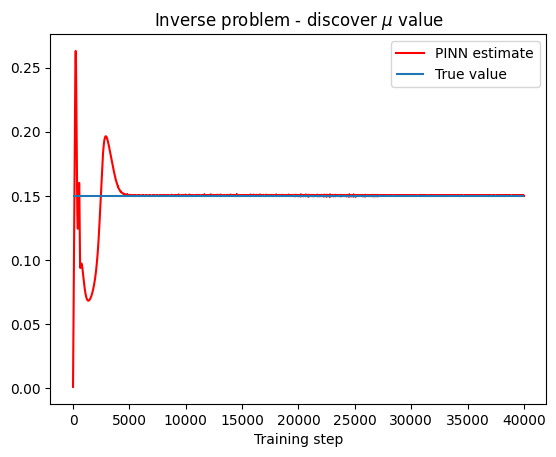}
  \caption{ (Left panel) The viscosity parameter value $\mu$ estimated by the PINNs solver during the training process, corresponding to the noisy data
  with a noise amplitude of $0.01$ (see previous figure). (Right panel) The viscosity parameter value $\mu$ estimated by the PINNs solver during the training process,
   corresponding to training data without noise.
         }
\label{fig13}
\end{figure}

\section{Discussion and conclusion}\label{sec:conclusion}

In this work, we have introduced the basic concepts of using neural networks in order to solve 2D partial integrate differential equations via the use of PINNs.

Benchmark tests on different second order PDEs are taken in order to illustrate the putting into practice. Very simple problems based on Laplace and
Poisson equations allowed to illustrate the use of vanilla-variant (soft boundary BCs with trainig data set) versus hard-variant (hard BCs via Lagaris-like
method). The latter problems are also used to show how different types of BCs can be imposed in PINNs solver.
More complex examples of PDEs taken from astrophysics are also considered. This is the case of Helmhotz equation for MHD modelling of magnetic
arcades equilibria in the solar corona. This is also the case of Grad-Shafranov equation for MHD modelling the structures of curved loops.
For the latter problems, PINNs solvers can easily deal with singularities (e.g. the singularity at $r=0$ for the physical Lane-Emden equation).
Note that, PINNs method can be also used to solve particularly a rather large set of $6$ partial differential equations representative of steady MHD model,
as successfully shown by Baty \& Vigon (2024) in the context of magnetic reconnection process.

The main advantage of PINNs when compared to traditional numerical integration methods is its numerical simplicity. Indeed, there is no need
of discretization, like in a finite-difference method for example. Additionally, The required number of collocation points in order to ensure the convergence of
the training is not so high. The second advantage is that, once trained the solution is instantaneously predicted on any other given grid (that can be
different from the collocation grid). This is not the case of traditional methods which generally require some additional interpolation from the computing grid.

Unfortunately, PINNs have however some drawbacks. The training process can be long for some problems. Indeed, this is related to 
the lack of general automatic procedure for a fine tuning of the hyper-parameters of the network in order to have an optimal
convergence during the training. The rules for the choice of the network architecture are not well  determined.
As concerns the accuracy, the precision is good but not as good compared to traditional schemes, and it can reveal insufficient for some problems.

Finally, PINNs can be particularly useful when solving parametric PDEs (as illustrated on parametric advection-diffusion ODE in this report). In a similar
way, PINNs can solve inverse problems for which a coefficient (or a term) is not known and must be discovered.

\subsection*{Acknowledgements}
Hubert Baty thanks, V. Vigon (IRMA and INRIA TONUS tream, Strasbourg) for stimulating discussions on PINNs technique, J. Pétri (Strasbourg observatory) for quoting the possibility to solve
Lane-Emden equations, and L. Baty (CERMICS, école des Ponts ENPC) for helping in the use of Python libraries and impromptu general discussions
on neural networks.

\section{Appendix}

We have plotted in Figures 35 and 36, the results obtained when solving Poisson problems (equations b-e, see main text) with Dirichlet and Neumann BCs
using vanilla-PINNs respectively.

\begin{figure}[h]
    \centering
    \includegraphics[width=0.8\textwidth]{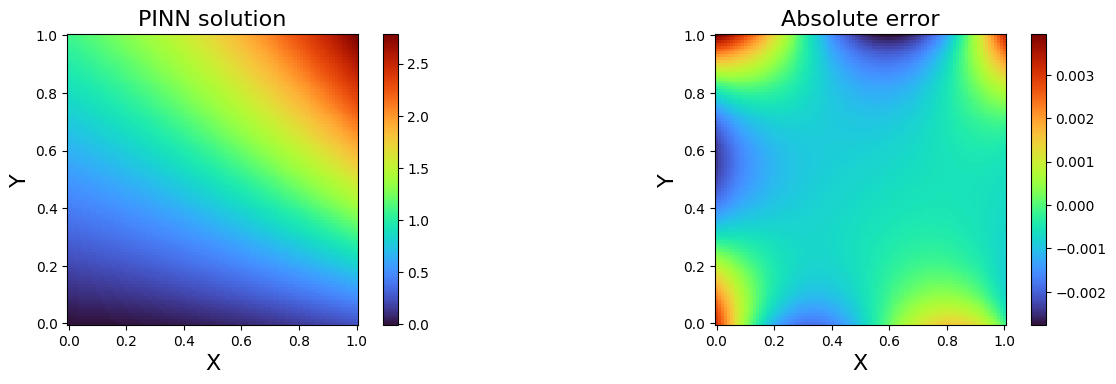}
    \includegraphics[width=0.8\textwidth]{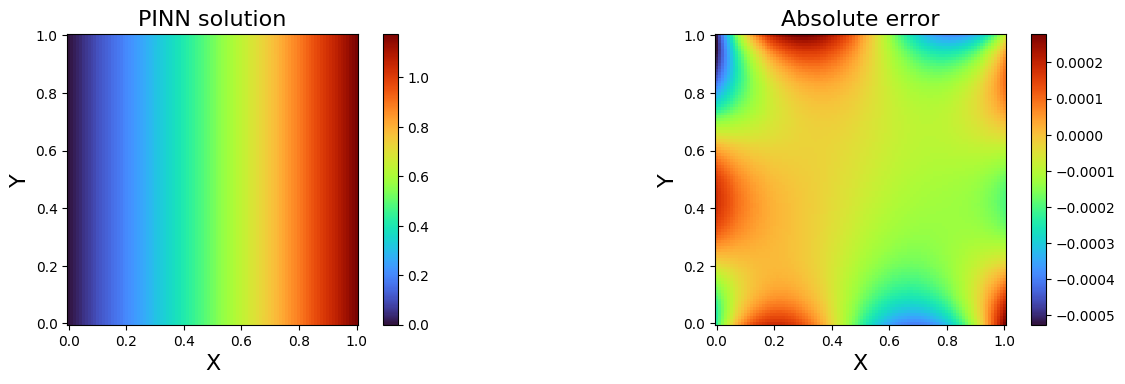}
    \includegraphics[width=0.8\textwidth]{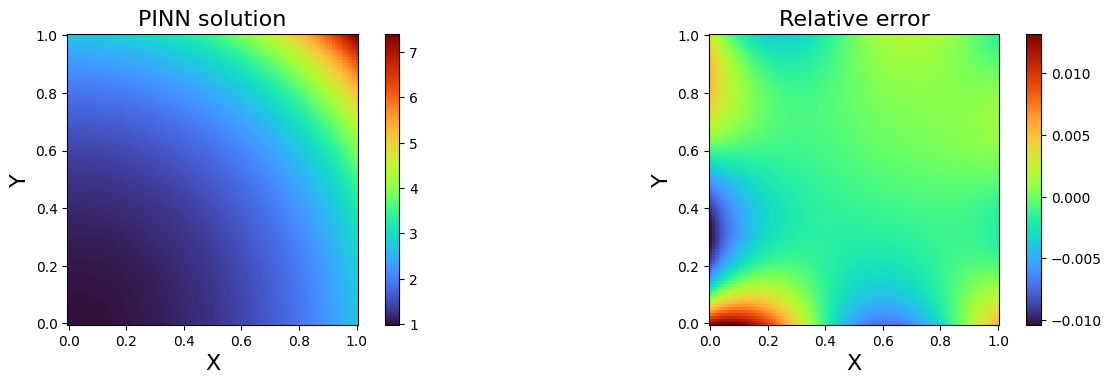}
     \includegraphics[width=0.8\textwidth]{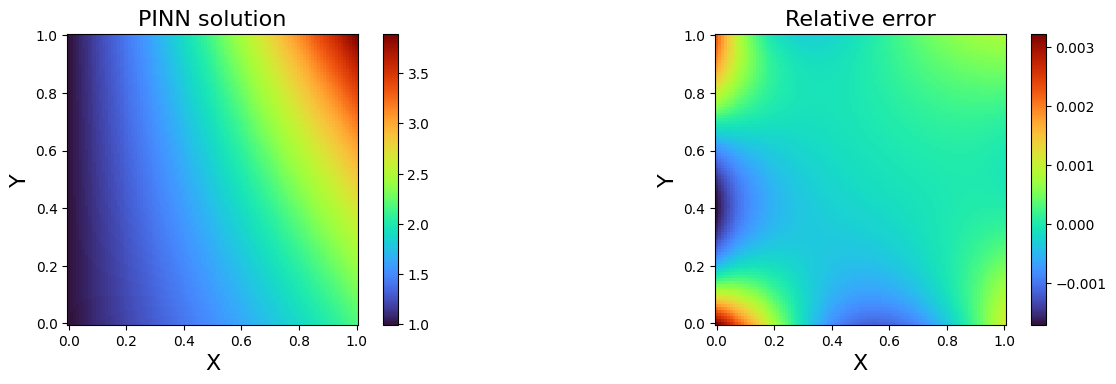}
  \caption{ PINNs solution and corresponding absolute/relative error distribution as colored iso-contours corresponding to Poisson problem with Dirichlet BCs
  using vanilla-PINNs. The four equations (b-e) examples are considered from top to bottom panels respectively (see text).
      }
\end{figure}

\begin{figure}[h]
    \centering
    \includegraphics[width=0.8\textwidth]{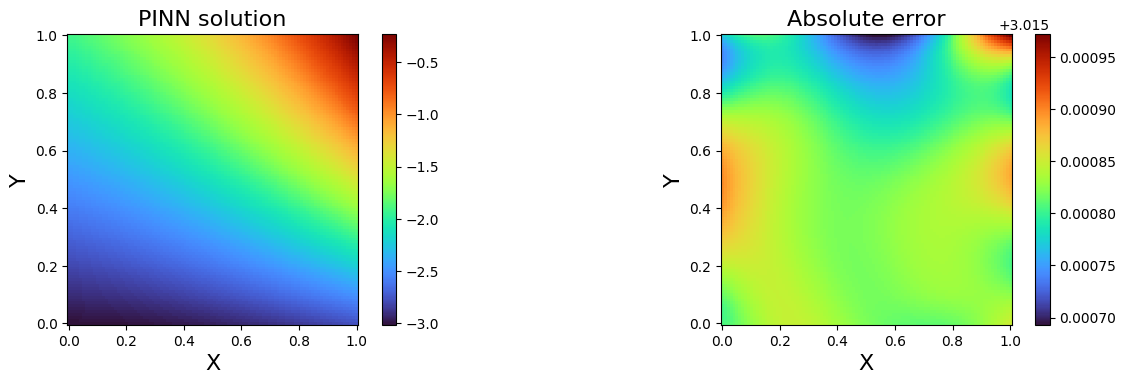}
    \includegraphics[width=0.8\textwidth]{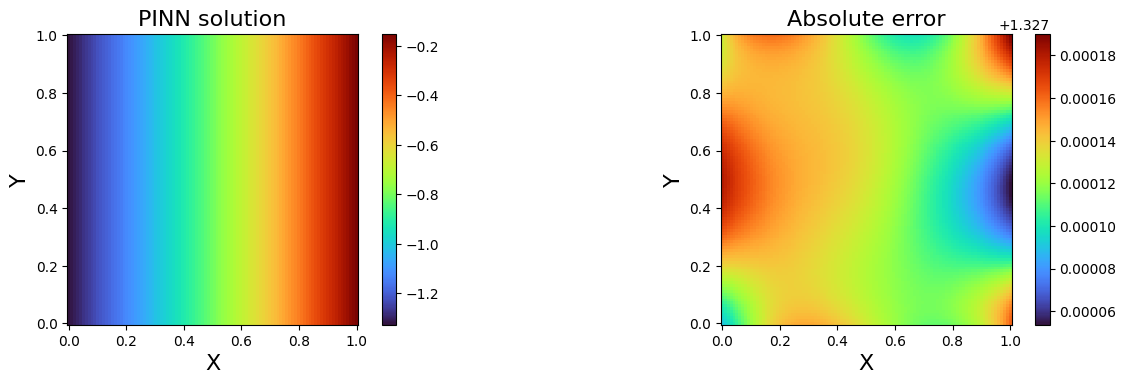}
  \caption{PINNs solution and corresponding absolute/relative error distribution as colored iso-contours corresponding to Poisson problem with Neumann BCs
  using vanilla-PINNs. The two equations (b-c) examples are considered from top to bottom panels respectively (see text).
      }
\end{figure}

\end{document}